\documentclass[aps,pre,floatfix,twocolumn,nofootinbib,showpacs]{revtex4} 
\usepackage{graphicx} \usepackage{amsmath} \usepackage{amssymb} 
\usepackage[sort&compress]{natbib}
\usepackage{graphicx}
\usepackage{amsmath}
\usepackage{amssymb}

\newcommand{\up}{\uparrow}

\newcommand{\Del}{\hat{\Omega}}

\newcommand{\del}{\delta w^2}

\newcommand{\Y}{\hat{\rho}^{1/2}\hat{\Omega}\hat{\rho}^{1/2}} 
\newcommand{\hX}{\hat{X}}
\newcommand{\q}{Q} 
\newcommand{\Pkl}{p(k|l,{\cal M}_\tau,{\cal M}_0)}
\newcommand{\Pl}{p(l|{\cal M}_0)} 
\newcommand{\Pkandl}{p(k,l|{\cal M}_0,{\cal M}_\tau)} 
\newcommand{\BEQ}{\begin{eqnarray}} 
\newcommand{\EEQ}{\end{eqnarray}} 
\newcommand{\BEA}{\begin{eqnarray}} 
\newcommand{\EEA}{\end{eqnarray}} 
 
\renewcommand{\d}{{\rm d}} 
 
\newcommand{\ep}{\varepsilon} 
 
\newcommand{\eps}{\varepsilon} 
 
\newcommand{\om}{\omega} 
\newcommand{\tr}{{\rm tr}}

\newcommand{\hrho}{\hat{\rho }} 
\newcommand{\hsigma}{\hat{\sigma }}

\newcommand{\hH}{\hat{H}} 
 
\newcommand{\hU}{\hat{U}} 
\newcommand{\W}{{\cal W}} 
 
\newcommand{\hPi}{\hat{\Pi}}

\renewcommand{\W}{{\cal W}} 
\renewcommand{\S}{{\cal S}} 
\newcommand{\E}{{\cal E}}

\newcommand{\half}{\frac{1}{2}}

\begin{document} 

\title
{The second law and fluctuations of work:\\
The case against quantum fluctuation theorems.}
\author{A.E. Allahverdyan$^{1,2)}$
and Th.M. Nieuwenhuizen$^{1)}$}
\affiliation{$^{1)}$ Institute for Theoretical Physics,
Valckenierstraat 65, 1018 XE Amsterdam, The Netherlands}
\affiliation{$^{2)}$Yerevan Physics Institute,
Alikhanian Brothers St. 2, Yerevan 375036, Armenia}

\begin{abstract} We study how Thomson's formulation of the second law:
no work is extracted from an equilibrium ensemble by a cyclic process,
emerges in the quantum situation through the averaging over fluctuations
of work.  The latter concept is carefully defined for an ensemble of
quantum systems interacting with macroscopic sources of work. The
approach is based on first splitting a mixed quantum ensemble into pure
subensembles, which according to quantum mechanics are
maximally complete and irreducible.  The splitting is done by
filtering the outcomes of a measurement process.  

A critical review is given of two other approaches to fluctuations of
work proposed in the literature.  It is shown that in contrast to
those ones, the present definition {\it i)} is consistent with the physical
meaning of the concept of work as mechanical energy lost by the
macroscopic sources, or, equivalently, as the average energy acquired
by the ensemble; {\it ii)} applies to an arbitrary non-equilibrium
state.  There is no direct generalization of the classical
work-fluctuation theorem to the proper quantum domain.  This implies
some non-classical scenarios for the emergence of the second law.
\end{abstract}

\pacs{PACS: 05.30.-d, 05.70.Ln}


\maketitle

\section{Introduction.}
\label{intro}

The second law was deduced in XIX'th century, and formulated for a
single closed system, in a way resembling the laws of
mechanics~\cite{epstein,tolman,landau,balian,fr,lindblad}. It was,
however, already the insight of Maxwell~\cite{scot,demon} and
Gibbs~\cite{gi} that this law has in fact a statistical character, and
refers to averages over an ensemble of identically prepared systems,
rather than to a single system.  This viewpoint became widely accepted
since the beginning of XX century, where first robust observations of
fluctuations were made \footnote{It was thus rather surprising to see recent
claims on ``violations of the second law'' \cite{evans} or ``transient
violations of the second law'' \cite{ritort} due to fluctuations; see in
this context our comment \cite{ANcontra}.}.  Together with theoretical
works of Boltzmann in kinetic theory of gases and of Smoluchowski,
Fokker, Planck and Einstein in physics of brownian motion, they formed a
consistent picture of the second law as emerging from microphysics
through averaging over fluctuations.  A detailed summary of this
activity is presented in the book by Epstein~\cite{epstein}, while
Tolman~\cite{tolman} discusses theoretical aspects of the situation.
Since then, the statistical understanding of the second law entered
several modern books of statistical physics and
thermodynamics~\cite{landau,balian}.  The current perspectives on the
classical and quantum brownian motion in the context of the second law
can be found in Refs.~\cite{sek1,AN,NA}. 

At the end of 1970's several groups independently gave a derivation of
Thomson's formulation of the second law
\cite{bk,bkaux,bkphysica,bassett,woron,lenard,thirring}: {\it no work
can be extracted from initially canonical equilibrium system by means of
a cyclic thermally isolated process}, starting directly from quantum or
classical Hamiltonian equations of motion. The very possibility of
getting such a straightforward thermodynamical result directly from
equations of motion is due to the fact that work is a transparent
quantity unambiguously defined both in and out of equilibrium for any
(quantum or classical) system interacting with external macroscopic work
sources \footnote{ These features of work are in contrast to those of
entropy, whose meaning is too closely tied to equilibrium states of
macroscopic bodies.}. As the main consequence, Thomson's formulation is
the only one which is valid both for finite and infinite systems which
do start in equilibrium, but can be driven arbitrary far from it by
external sources (see \cite{an} and sections \ref{fufu} and
\ref{conclusion} for more details). 

The standard understanding of the second law and fluctuations is based
on Einstein's formula relating entropy to the probability of a
fluctuation around equilibrium ~\cite{epstein,tolman,landau,fr}.  This
suffices for the purposes of near-equilibrium thermodynamics of
macroscopic bodies, in particular, because all the formulations of the
second law are equivalent for them and entropy is defined
unambiguously. In the more general case of finite systems and/or
systems driven strongly out of their initial equilibrium, relations
between the second law and fluctuations ought to be studied anew for
each meaningful formulation of the law in separate.  The conclusions may
differ from case to case and from formulation to
formulation \cite{NA,an}. 

The purpose of the present paper is to understand how Thomson's
formulation of the second law in the quantum situation emerges through
the averaging over fluctuations of the equilibrium ensemble. More
specifically, if the (average) work done on the initially equilibrium
ensemble during a cyclic process is always non-negative, what are
fluctuations of this work, and how do they behave?  There are definite
answers to these questions in the classical situation: the definition of
fluctuations of work is straightforward, and a model-independent
information on them is given by an equality first derived by Bochkov and
Kuzovlev in 1978~\cite{bk} (BK equality). Later on, this equality,
sometimes also called work-fluctuation theorem, was extended to
non-cyclic processes \cite{jar}, and has undergone various
generalizations \footnote{\label{gengen} A rather complete account of
various generalizations of the classical work-fluctuation theorem, as
well as its relation with other fluctuation theorems, e.g., those
describing entropy production, is given in Refs.~\cite{crooks,ma}.
Local versions of the fluctuation theorems are also discussed there.}.
The basic messages of the classical BK equality are recollected and
reviewed in section \ref{messages}. 

While all these developments concern the classical situation, a number
of recent works is devoted to quantum extensions of BK equality
\cite{kurchan,yukawa,tasaki,mukamel,maes,sasaki}. The first definition
of fluctuations of work in the quantum situation and a quantum
extension of BK equality was actually proposed by Bochkov and Kuzovlev
themselves \cite{bk,bkphysica}. It is based on constructing an
operator in the Heisenberg representation, associating it
with an operator of work, and thus treating work as an ordinary
quantum mechanical observable pertaining to the system and not to
the work-source.

Another extension was initiated by Kurchan \cite{kurchan}, based on
two-time measurements of energy. As implied by the standard treatment of
quantum measurements, this second approach is closely tied to
Schr\"odinger representation. 

There are therefore two different approaches to the definition of
fluctuations of work and to quantum extensions of BK equality; both
of them attracted attention recently
\cite{yukawa,tasaki,mukamel,maes,sasaki}, and are reviewed below in
section \ref{others}. The fact that in the quantum situation these
approaches are different is calling for attention. The difference in
viewpoints is not
completely unexpected, since the work as it appears in statistical
thermodynamics \cite{tolman,landau,balian,fr} is an essentially
classical quantity (mechanical energy transferred from a classical
source of work).

Our objective is to propose a third definition of fluctuations of
work, which is motivated by the fundamental physics of quantum 
(sub)ensembles.  The
definition is guided by the following observation.  Since the usual
work is now presented as an average of a random quantity | for the
moment we leave unspecified whether this is a random classical
quantity or an operator | it is natural to require the following two
conditions on its (fluctuating) realizations and on its average:

\begin{itemize}

\item Once the average work is unambiguously defined for any quantum
system starting in an arbitrary initial state and interacting with a
macroscopic source of work, the same should hold for fluctuations of
work.  In particular, one {\it cannot} restrict the general definition
to (initially) equilibrium states of the system.  This is relevant,
since one of the chapters in statistical thermodynamics deals with work
extraction from non-equilibrium systems~\cite{landau,abn}, and one should,
of course, be able to define fluctuations of work in this most general
situation. 

\item Realizations of the random quantity work should have the same
physical meaning of mechanical (high-graded) energy as the usual
(average) work. In particular, if one happens to extract some work
from a single realization, one should be able | at least in principle
| to use precisely this amount for the standard purposes, e.g. for
driving motors. (Basic features of work are recalled below in section
\ref{workclass}.)

\end{itemize}

Both these conditions are satisfied by the classical definition, and
to our opinion without them the very program of studying the
emergence of the second law in the quantum situation becomes ill
defined.

It appears to the present authors that, as we discuss below in section
\ref{others}, neither of the existing two quantum approaches |in the
way they stand presently| can be viewed as providing a proper
definition of fluctuations of work in the quantum situation.  Both
approaches fail out of equilibrium (no first condition), while even
for an initially equilibrium state it is not clear that the
second condition is satisfied. Neither these points were discussed in
papers which support those definitions; see e.g.,
\cite{bk,bkphysica,yukawa,sasaki,kurchan,mukamel,maes}.

These are the reasons to introduce in the present paper a third
approach to quantum fluctuations of work, which will satisfy the above
two conditions.  Our approach starts with explicitly respecting the
first condition, that is, always defining realizations of (the random
quantity) work as an average energy given off by the macroscopic
source of work.  If the corresponding ensemble of physical systems
already consists of subensembles, non-trivial realizations can be
defined via the average energy exchange of each subensemble with the
source. (As with any exchange process, this is operationally
characterized by measurements at two different times.)  For a
classical ensemble each single member completely characterizes a
subensemble, and the classical definition of fluctuations of work
follows naturally.  In contrast, a quantum equilibrium state is
described by a homogeneous quantum ensemble, the gibbsian, which by
itself does not consist of subensembles. This prevents us to proceed
as such. First, an inhomogeneous quantum ensemble has to be prepared
from this homogeneous one, a task accomplished via a selective quantum
measurement. The obtained structure of subensembles does depend on the
type of measurement, and as a consequence the resulting fluctuations
of work in the quantum situation appear to be context-dependent
(contextual). Still some relevant characteristics of these
fluctuations can be context-independent, as seen below.

As the main result, the second law in Thomson's formulation | whose
statement reads in the same way both in quantum and classics | has in
those two situations rather different scenario of emergence.  The
basic qualitative difference is that in contrast to classics, the
fluctuations of work in the proper quantum situation are not
controlled by any direct analog of BK equality. More specifically, in
classics the structure of work as a random quantity is such that there
have to be realizations which provide work (i.e., which are active).
In the quantum case, however, there need not be any active realization.

In this paper we have taken the simplest approach which allows to
study Thomson's formulation of the second law and fluctuations of
work, that is, we consider a finite quantum or classical system
interacting with external sources of work.  The restriction to finite
| though possibly large | systems is at any rate natural for studying
fluctuations.  The explicit presence of thermal baths, as well as
details about the thermodynamical limit for the system are left for
future studies.

The paper is intended to be self-consistent and is organized as follows.
In section \ref{class} we recall the definition of fluctuations of
work in the classical situation and review the BK equality and its
consequences.  In section \ref{quant} we present the definition of
fluctuations of work in the quantum situation. The dispersion of work
is studied in section \ref{dispersion}. In section \ref{no} we show
that fluctuations of work in the quantum situation are not controlled
by any direct analog of the classical BK equality.  An anti-classical
scenario for the emergence of the second law in Thomson's formulation
is described in section \ref{dispersion}.  In section \ref{others} we
make comparison with the two known approaches on fluctuations of work
in the quantum situation.  These approaches offer different extensions
of classical BK equality. We do not intend to imply that these
approaches do not have a physical meaning or that they cannot
be useful for their own sake. We only state that |in the way they
stand presently| they do not describe fluctuations of work in the
proper quantum situation.  We close with a discussion. Some details are
worked out in appendices.

\section{Classical fluctuations of work and BK equality.}
\label{class}

\subsection{The setup.}

Consider an ensemble $\E$ of identical classical systems $\S$ which are
thermally isolated \cite{landau,balian}: they move according to their
own dynamics and interact with an external macroscopic work source $\W$.
This interaction is described via time-dependence of some parameters
$R(t)=\{R_1(t),R_2(t),...\}$ of the system's Hamiltonian
$H(t)=H\{R(t)\}$; see Refs.~\cite{landau,balian}. 

The parameters move along a certain trajectory $R(t)$ which at some
initial time $t=0$ starts from $R(0)$, and ends at $R(\tau)$ at the
final time $t=\tau$.  Cyclic thermally isolated processes are defined by
$R(0)=R(\tau)$ and thus \BEA H\{R(\tau)\}=H\{R(0)\}\equiv H.\EEA At
the initial time the ensemble is in equilibrium, that is, the common
probability distribution ${\cal P}(x,p;t=0)\equiv {\cal P}(x,p)$ of
all its canonically conjugated coordinates $x=(x_1,...,x_n)$ 
and momenta $p=(p_1,...,p_n)$ is
given by the Gibbs distribution with the initial Hamiltonian $H(x,p)$
and temperature $T=1/\beta\geq 0$: 
\BEA
\label{00}
{\cal P}(x,p)
=\frac{e^{-\beta H(x,p)}}{Z},\quad
Z=\int\d x\,\d p \,e^{-\beta H(x,p)}.
\EEA
This equilibrium distribution can be prepared by means of a thermal bath
coupled with the system $\S$ for $t<0$.
It is assumed that 
for times $0\leq t\leq\tau$ the system $\S$ is decoupled from the bath |an
alternative assumption would be that its coupling to the bath is so weak
that it can be neglected in the considered time-interval|
and the evolution of the ensemble is described by
the Liouville equation for ${\cal P}(x,p;t)$:
\BEA
\label{evol}
\partial_t{\cal P}(x,p;t)=
\frac{\partial H(x,p,t)}{\partial p}\frac{\partial {\cal P}
(x,p;t)}{\partial x}\nonumber\\
-\frac{\partial H(x,p,t)}{\partial x}\frac{\partial {\cal P}
(x,p;t)}{\partial p}.
\EEA

\subsection{Work}
\label{workclass}

In statistical thermodynamics there are two alternative definitions of
work \cite{tolman,landau,balian,lindblad,perrot}. Both are necessary
for the proper understanding of its physical meaning 
\cite{balian,perrot,abnpreparation}. The first reads
\begin{itemize}

\item The work $W$ is the average energy gained by $\S$ during a 
thermally isolated system-work-source
interaction with $\W$~\cite{landau,balian}:
\end{itemize}
\BEA 
\label{work-1}
W=
\int\d x\,\d p\,[\,{\cal P}(x,p;\tau)\,H(x,p;\tau)
\nonumber\\
- {\cal P}(x,p)\,H(x,p;0)\,].
\EEA

Due to conservation of energy, $W$ is equal to the average energy lost
by the work-source $\W$. 
This definition was (implicitly) proposed by
Caratheodory~\cite{perrot}.  A concise history of various definitions
of work is given in \cite{am1}, while various perspectives of work in
classical mechanics are reviewed in \cite{amleff}.

For cyclic processes Eq.~(\ref{work-1}) takes a simpler form
\BEA W
=\int\d x\,\d p\,[{\cal P}(x,p;\tau)-{\cal P}(x,p)]\,H(x,p;0).
\label{work0} 
\EEA

There is a second, alternative definition going back to Gibbs and Planck
\cite{perrot,am1}:

\begin{itemize}

\item The minus work $-W$ is the energy transferred to the work-source
$\W$.  Its distinguishing feature with respect to other forms of
energy is that it can, in principle, be transferred with 100 \%
efficiency to other work sources via interactions of the
system-work-source type. In particular, it can be retransferred to
collective degrees of freedom that perform {\it classical
deterministic} motion generated by a suitable Hamiltonian.  These
degrees of freedom are thus purely mechanical and serve as prototypes
of macroscopic mechanical devices (such as a motor, piston, turbine,
etc.). For them the differential work can be calculated in the usual
way of ordinary mechanics, that is, multiplying the external force
by the corresponding displacement \cite{balian}.

\end{itemize}

Both these definitions of work are expected to be
equivalent at least for sufficiently ideal work sources
\cite{balian,perrot,abnpreparation}.

Yet we mention for completeness another, equivalent
formula for the work $W$; the integral of the rate of 
energy change: 
\BEA
\label{vko}
W=\int_{0}^{\tau}\d t
\int\d x\,\d p \,
{\cal P}(x,p;t)\,\frac{\partial H(x,p;t)}{\partial t}.  
\EEA
To get from
here to Eq.~(\ref{work0}) one performs integration by parts, uses the
standard boundary conditions, that is ${\cal P}(x,p;t)$ decays for
$x\to\pm\infty$ or $p\to\pm\infty$ and employs Eq.~(\ref{evol}).  
This formula for $W$ is more general and can be applied to processes
that involve explicit thermal baths.

\subsection{Fluctuations of work}
\label{fufu}

Though the ensemble $\E$ is described by the probability
distribution ${\cal P}(x,p)$, each single system $\S$ from this
ensemble has at a given moment of time explicit values for all its
dynamical variables. These values may vary from one single system to
another due to the distribution of initial conditions.

Each single member of the ensemble is then coupled to the external
source of work that realizes on it a unique thermally isolated process
(the same for all members).
In other words, the same parameters $R(t)$ of the
Hamiltonian are varied in the same way for each member.  The
motion of the single system is described by
Eq.~(\ref{evol}) with now ${\cal P}(x,p;t)$ being a product of two
delta-functions $\delta(x-x(t))\,\delta(p-p(t))$, which are
probability densities concentrated at the solutions of the canonical
equations of motion:
\BEA
\label{canon}
\dot{p}=-{\partial_xH(x, p;t)},\qquad \dot{x}={\partial_p H(x, p;t)}.
\EEA 
The trajectories generated by (\ref{canon}), together
with their initial conditions distributed according to
Eq.~(\ref{00}), serve as {\it realizations} of the random process given by
Eq.~(\ref{evol}).  

The work $w(x,p)$ exchanged in each thermally isolated
process can then be calculated consistently with Eq.~(\ref{work0}):
\BEA
\label{8-}
w(x,p)=H(x(\tau), p(\tau);\tau)-H(x,p)\\
=H(x(\tau), p(\tau))-H(x,p),
\label{fluwork}
\EEA
where $H(x(\tau), p(\tau);\tau)$ is the value of the Hamiltonian on
the trajectory that started at $t=0$ from $(x,p)$, with $x(\tau)$ and
$p(\tau)$ being the corresponding solutions of (\ref{canon}).  This
work can be observed as the energy decrease of the mechanical degree
of freedom of the macroscopic work-source, or alternatively via energy
increase of the system $\S$. In this latter scenario the energy of $\S$
has to be measured two times, at the moments $t=0$ and $t=\tau$.

The so defined work $w(x,p)$ for a single system is a random quantity,
since it varies from one single system to another. It can be positive or negative.
Its probability
distribution $P(w)$ is determined by ${\cal P}(x,p)$, since this is
the probability by which each single system enters in the ensemble:
\BEA
\label{gaza}
P(w)=\int \d x\,\d p\,
{\cal P}(x,p)\delta(w-w(x,p)).
\EEA

There being used no special features of the initial equilibrium
distribution function, the same definition for the work in a
single realization can be given for any initial ensemble.  

It is seen that the two desired conditions for fluctuations of work
formulated in section \ref{intro} are naturally satisfied: the initial 
distribution may be arbitrary and ``work for a single realization'' 
has the same physical meaning as average work.

\subsection{Derivation of BK equality.}

One now derives BK equality in the classical situation for a closed cycle,
\cite{bk,bkphysica,jar}:
\BEA
\label{gort}
\langle e^{-\beta w}\rangle&\equiv&
\int \d w\,P(w)\,e^{-\beta w}\\&=&
\int \d x\,\d p\,{\cal P}(x,p;0)\,
e^{-\beta w(x,p)}\nonumber\\&=&
\frac{1}{Z(0)}\int\d x\,\d p\,
e^{-\beta H(x, p)-\beta w(x,p)}\nonumber\\&=&
\frac{1}{Z(0)}\int\d x\,\d p\,
e^{-\beta H(x(\tau), p(\tau);\tau)}\nonumber\\&=&
\frac{1}{Z(0)}\int\d x(\tau)\,\d p(\tau)\,
e^{-\beta H(x(\tau), p(\tau);\tau)}\nonumber\\&=&\frac{Z(\tau)}{Z(0)}
=1,
\label{dodosh}
\EEA
where we used Liouville' theorem $\d x\,\d p=\d x(\tau)\,\d p(\tau)$ and
Eqs.~(\ref{00}, \ref{fluwork}, \ref{gaza}). 
The last equality in (\ref{dodosh})
is due to the assumed cyclic feature of the process.

\subsection{Qualitative messages of the BK equality.}
\label{messages}

The
BK equality is by itself an exact mathematical relation. 
Several important qualitative results
can be deduced from it: 

\paragraph{The second law.}
\label{message1}

As the exponential function is convex, one gets directly $1=\langle
e^{-\beta w}\rangle\geq e^{-\beta\langle w\rangle}$, and then
$W=\langle w\rangle\geq 0$, which is the statement of the second law
in Thomson's formulation: no work can be extracted from an equilibrium
system by means of a cyclic process. This formulation of the second law
is well-known and has an independent and more general derivation both
in the classical and the quantum situation
\cite{bk,woron,lenard,bassett,thirring,ANphysicaA}.

\paragraph{Active realizations.}
\label{message2}

To satisfy
$1=\langle e^{-\beta w}\rangle$ directly leads to the
following observation: for any cyclic thermally isolated process there are
realizations which are active, that is, for which 
work is extracted after the process: $w(x,p)<0$. The relative weight
of such active realizations can be estimated via the Cauchy inequality:
\BEA
1=\left(\int \d x\,\d p\,\sqrt{{\cal P}(x,p)}\,\sqrt{{\cal P}(x,p)\,}
e^{-\beta w(x,p)}
\right)^2\nonumber\\
\leq \int \d x\,\d p\,{\cal P}(x,p)
\int \d x\,\d p\,{\cal P}(x,p)
\,e^{-2\beta w(x,p)},
\EEA
which can be written as
\BEA
\label{tutmos1}
\langle e^{-2\beta w}\rangle\geq 1.
\EEA
A stronger relation can be obtained with help of
a generalization of the Cauchy inequality, described in Appendix 
\ref{mitri}. It reads:
\BEA
\label{tutmos2}
\langle e^{-2\beta w}\rangle\geq 
1+\frac{\left[\langle \,[\,f-\langle f\rangle\,]\,
e^{-\beta w}\,\rangle\right]^2
}{\langle \,[\,f-\langle f\rangle\,]^2\,\rangle}
> 1,
\EEA
where $f(x,p)$ is an arbitrary integrable function in the phase-space,
and where 
\BEA
\langle f\rangle\equiv \int \d x\,\d p\,{\cal
P}(x,p)\,f(x,p).  
\EEA
Eq.~(\ref{tutmos2}) is stronger than (\ref{tutmos1}), since now
$\langle e^{-2\beta w}\rangle$ is shown to be strictly larger than
$1$.  Inequalities (\ref{tutmos1}, \ref{tutmos2}) allow to understand
how relevant the active realizations are with respect to both their
probability and the amount of extracted work.

\paragraph{Dispersion of work.}
\label{message3}

For sufficiently high temperatures one can make a cumulant
expansion:
\BEA
1=\exp[-\beta \langle w\rangle+\frac{\beta^2}{2} (\langle w^2\rangle
-\langle w\rangle^2)+...]
\EEA
which shows that for sufficiently high temperatures the ratio of the
dispersion of work $\langle w^2\rangle-\langle w\rangle^2$ and its
average increases with temperature:
\BEA
\frac{\langle w^2\rangle-\langle w\rangle^2}{\langle w\rangle}
=2T.
\EEA

A detailed survey of various cumulant expansion-based results 
derivable from BK equality is contained in 
Refs.~\cite{bk,bkaux,bkphysica}. 

\subsection{Non-cyclic processes.}
\label{noncyclic}

For non-cyclic processes there is an analog of equality (\ref{dodosh}),
which is derived in similar way with the conclusion \cite{jar}:
$\langle\,e^{-\beta w}\,\rangle=e^{-\beta (F(\tau)-F(0))}$, where
$F=-T\ln Z$ is the corresponding free energy.  This relation allows to
calculate differences of free energy via (non-equilibrium) measurements
of work. This generalized equality is not directly relevant for our
present purposes, because here we are interested by the second law in
Thomson's formulation which refers to cyclic processes. 

\section{Quantum ensembles and 
the definition of fluctuations of work}
\label{quant}

\subsection{The setup.}

The quantum setup for studying thermally isolated processes is a
straightforward extension of the classical one.  
(We denote all operators by a hat.)

An ensemble $\E$ of identically prepared quantum systems 
$\S$ is described at $t=0$ by 
a density matrix $\hrho(0)=\hrho$.
The eigenresolutions of $\hrho$
and of the Hamiltonian $\hH$ read:
\BEA
\label{resolrho}
\hrho&=&\sum_{k=1}^n
p_k|p_k\rangle\langle p_k|, \\
\label{resolH}
\hH&=&\sum_{k=1}^n
\eps_k|\eps_k\rangle\langle\eps_k|, 
\EEA
where 
$\{|\eps_k\rangle\}_{k=1}^n$ and $\{|p_k\rangle\}_{k=1}^n$ 
with $\langle\eps_k|\eps_l\rangle=\langle p_k|p_l\rangle
=\delta_{kl}$ are the eigenvectors of 
$\hH$ and $\hrho$, respectively, which
form bases in the $n$-dimensional Hilbert space ${\cal H}$, and where 
$\eps_k$ and $p_k$ are the corresponding eigenvalues.

Frequently, but not always, we will consider initially Gibbsian states:
\BEA 
\label{gibbs} 
\hrho(0)=\hrho=\frac{e^{-\beta \hH}}{Z},
\quad 
Z={\rm tr}\,e^{-\beta \hH}, \\
\label{pk}
p_k=\frac{e^{-\beta \eps_k}}{\sum_{k=1}^ne^{-\beta \eps_k}},\quad
|p_k\rangle=|\ep_k\rangle,\quad k=1,..,n,
\EEA 
where $T=1/\beta\geq 0$ is
the temperature of the ensemble.
We shall order the eigenvalues of $\hH$ as
\BEA
\label{dunaj}
\ep_1\leq \ep_2\leq...\leq \ep_n.
\EEA
Then according to (\ref{pk}), the eigenvalues of $\hrho$ will be
ordered as
\BEA
p_1\geq p_2\geq...\geq p_n>0.
\EEA
For the Gibbsian density matrix all eigenvalues are positive.

Analogously to the classical case, the Gibbsian state (\ref{gibbs}) is
prepared for $t<0$ by letting $\S$ to interact with a macroscopic
thermal bath, and then decoupling it from the bath, so that the
interaction is absent for $t>0$. There is, however, a relevant
difference between quantum and classical: in the quantum situation
the coupling of $\S$ with
the bath has to be weak for the stationary state of $\S$ to be
Gibbsian \footnote{\label{mut} Due to weak coupling to the bath, the
energy costs for switching the interaction on and off become
negligible. This holds both in the quantum and the classical situation.}. A
detailed analysis of this and similar differences between the
emergence of Gibbs distribution in quantum and classical situations is
presented in \cite{AN,NA}.

At $t=0$ $\S$ starts to interact with an external macroscopic work source $\W$.
The resulting evolution of $\S$ is generated by (an effective)
Hamiltonian $\hH\{R(t)\}$, which is time-dependent via classical
(c-number) parameters $R(t)$. The evolution of $\S$ is thus unitary
and has the same general features of reversibility as the dynamics of
a completely isolated $\S$.  It is well known that in general a
Hamiltonian evolution of the complete system $\S+\W$ does not reduce
to a Hamiltonian evolution for the state of $\S$. However, in the
present case this is achieved owing to the {\it macroscopic} character
of $\W$, as discussed in \cite{balian}.

A cyclic process at the moment $t=\tau$ is defined in the same
way as in classics, that is, via $R(\tau)=R(0)$, leading to
\BEA
\hH(\tau)=\hH(0)=\hH.
\EEA

The Hamiltonian $\hH(t)$ generates a unitary evolution:
\BEA 
\label{evolution} 
i\hbar\frac{\d}{\d t}{\hrho}(t)=[\,\hH(t),\hrho(t)\,],\\
\label{ddd}
\hrho(t)=\hU_t\,\hrho(0)\,\hU_t^\dagger, \\
\hU_t=\overleftarrow{\exp}\left[-\frac{i}{\hbar}\int_{0}^{t}\d s\,
\hH(s)\right],
\label{unita}
\EEA 
where $\overleftarrow{\exp}$ and $\overrightarrow{\exp}$ denote
time-ordered and time-anti-ordered exponents, respectively.

\subsection{Work.}
\label{workquant}

The whole discussion in section \ref{workclass} directly applies in
the quantum situation, except that $\S$ is now a quantum system, and
Eqs.~(\ref{work-1}, \ref{work0}) should be substituted by their
quantum analogs (i.e., ${\cal P}\to \hrho$, $H\to \hH$ and $\int\d x\,\d
p\to\,{\rm tr}$).
More specifically, the work $W$ done by the
external source $\W$ is identified with the average energy gained by
the ensemble \cite{landau,balian}
\BEA 
W={\rm tr}[\hrho(\tau)\,\hH- 
\hrho\,\hH]={\rm tr}\,\hrho\,\Del,
\label{quantwork} 
\EEA 
where we denoted 
\BEA
\label{deldef}
\Del\equiv\hU^\dagger_\tau\,\hH(\tau)\,\hU_\tau
-\hH=\hU^\dagger_\tau\,\hH\,\hU_\tau-\hH.
\EEA
Here $\hU^\dagger_\tau\,\hH(\tau)\,\hU_\tau$ is the Hamiltonian
operator in the Heisenberg representation at the end-time $\tau$
of the cyclic process.
The operator $\Del$ is sometimes called `operator of work'
\cite{lindblad,bk,yukawa}.  We shall show, however, in section
\ref{noopofwork} that it is not clear whether it fulfill all criteria
to deserve this identification.  Moreover, the much weaker interpretation
of $\Del$ |by analogy to the classical expression (\ref{8-})| as
``energy difference operator in the Heisenberg representation'' is
also incorrect in general; see section
\ref{noopofwork}. In our approach $\Del$ will always appear inside averages
over density matrices, so we do need any more particular interpretation
of $\Del$; it will only enter the definition of work (\ref{quantwork}).

The remarks we made after Eq.~(\ref{work0}) for the classical
situation are valid in the quantum case as well. $W$ is equal to the
average energy decrease of the work source $\W$. This is a classical,
mechanical energy which can transferred with 100\% efficiency to other
work-source, and, in particular, it can transferred to another mechanical
degree of freedom performing classical deterministic motion.  In that
respect both the classical and quantum definitions are consistent and
can be indistinguishable from the viewpoint of this mechanical
degree. This property is the underlying reason
why phenomenological thermodynamics, where not any
(quantum or classical) identification of $\S$ is given, can exist.

The work is typically observed via suitable (classical) measurements
done on the work source, or, alternatively, by measuring the initial
and final average energies on the ensemble ${\cal E}$.  Both these
ways are routinely employed in practice, e.g., in NMR/ESR physics,
where the system $\S$ corresponds to spin-$\frac{1}{2}$ under influence
of external magnetic fields \cite{abo}. 

Finally, the quantum analog of formula (\ref{vko}) reads:
$W=\int_0^\tau\d t\, {\rm tr}\left[\hrho(t)\, \frac{\d \hH(t)}{\d
t}\right]$, and Eq.~(\ref{quantwork}) can be recovered from this
formula upon integration by parts and using (\ref{evolution}).

\subsection{Quantum ensembles.}

The definition of fluctuations of work in the classical situation was
based on the distinction between classical ensemble of systems
described by a probability distribution versus a single member of that
ensemble. It should not be surprising that fluctuations of work in
the quantum situation are closely tied to the meaning of what is a quantum
ensemble.

Thus, for our further purposes we need an account of various features
of quantum ensembles and their differences with respect to the
classical ones. There are several sources in literature
\cite{elsasser,park,espagnat,peres,willi,petr} where this type of
questions is studied with special attention \footnote{ Though the
theory of quantum ensembles is almost as old as quantum mechanics
itself, it still attracts lively discussions; see e.g.
\cite{cohen,terno,kirkespagnat}. It is interesting to note the basic
differences between classical and quantum ensembles were correctly
understood by Elsasser as early as in 1937 \cite{elsasser}.}.

\subsubsection{Statistical interpretation of quantum mechanics.}

Within the standard quantum mechanics a quantum `state' is
described by a density matrix $\hat\rho$.  Any state, including a pure
state $|\psi\rangle\langle\psi|$, describes an ensemble of 
identically prepared systems
For instance, in an ideal Stern-Gerlach
experiment all particles of the upper beam together are described by
the wavefunction $|\!\up\rangle$ or the pure density matix
$|\!\up\rangle\langle\up\!|$.  The description is optimal, in the
sense that all particles have $\sigma_z=+1$, but incomplete in the sense
that their $\sigma_x$ and $\sigma_y$ 
are unknown: upon measuring either of them,
one will get $\pm 1$ with equal probabilities.

\subsubsection{Homogeneous ensembles.}

In general, a density matrix $\hrho$ can be applied to describe two
types of quantum ensembles, {\it homogeneous} and {\it inhomogeneous}.

For a homogeneous ensemble $\E(\hrho)$ only the density matrix $\hrho$
is given and no further specification is made about a single system
$\S$ from that ensemble.  A typical example is an ensemble prepared by
thermalization, that is, by letting each single system $\S$ to
interact weakly with an equilibrium thermal bath, and waiting
sufficiently long till the equilibrium state of $\S$ is established.

Let us study the defining feature of homogeneous ensembles in more
details. We start by comparing them to classical ensembles.
In the classical situation, the description of an
ensemble by means of a probability distribution
still implies that each single system
has definite values for {\it all} its variables.  For a homogeneous quantum
ensemble $\E(\hrho)$, only those observables (hermitian operators living in
the Hilbert space ${\cal H}$) 
$\hat{A}$ that are dispersionless on $\E(\hrho)$,
\BEA
\label{dhol}
\left[{\rm tr}\left(\hat{A}\, \hrho
\right)\right]^2= {\rm tr}\left(\hat{A}^2\, \hrho
\right),
\EEA
can be said to have definite values for all single systems $\S$ from
$\E(\hrho)$ .  Indeed, it is shown in Appendix \ref{f7} that
dispersionless observables satisfy
\BEA
\label{duduk}
\hat{A}\,\hrho=\alpha\,\hrho,
\EEA
where $\alpha$ is a c-number. This implies 
\BEA
{\rm tr}\,(\,\hat{A}^m\,\hrho\,)=
\left[{\rm tr}\,\hat{A}\,\hrho\right]^m, 
\quad m=0,1,2,3...,
\EEA
and the above statement follows.
For a pure state $\hrho= |\psi\rangle\langle\psi|$, we return from
(\ref{duduk}) to the standard notion of $|\psi\rangle$ being an
eigenstate of $\hat{A}$.

Any other, non-dispersionless observable $\hat{B}$ | even if it
commutes with the density matrix $\hrho$ | does not have a definite
value in a single system $\S$ from $\E(\hrho)$. It is true that for
$[\hrho,\hat{B}]=0$, $\E(\hrho)$ can be prepared by mixing \footnote{
Mixing ensembles $\E(\hrho_1)$ and $\E(\hrho_2)$ with probabilities
$p_1$ and $p_2$, respectively, means that one
throws a dice with probabilities of outcomes equal to $p_1$ and
$p_2$, and depending on the outcome one picks up a system from
$\E(\hrho_1)$ or $\E(\hrho_2)$, keeping no information on where
the system came from. Alternatively, one can 
join together
$Np_1$ systems from $\E(\hrho_1)$ and $Np_2$ systems from
$\E(\hrho_2)$ ($N\gg 1$), so that no information information is kept
on where a single system came from. Then any subensemble
of $M$ systems ($N\gg M$) is described by
the density matrix $\hrho=p_1\,\hrho_1+p_2\,\hrho_2$. Note that the
restriction $N\gg M$ is important, see, e.g., \cite{terno}, and some 
confusion arose in literature for not taking it into account.
} pure states ensembles
$\{\,\E(|p_k\rangle\langle p_k|)\,\}_{k=1}^n$ with probabilities
$\{p_k\}_{k=1}^n$, where $\{\,|p_k\rangle\,\}_{k=1}^n$ and
$\{p_k\}_{k=1}^n$ are, respectively, the common eigenvectors of
$\hrho$ and $\hat{B}$ and the eigenvalues of $\hrho$.
If $\E(\rho)$ is {\it known} to be prepared in such a way, then
$\hat{B}$ has indeed definite values for each single member of
$\E$. However, in general this need not apply, since there are 
(infinitely) many other ways to prepare the same ensemble $\E(\hrho)$
via mixing $N$ subensembles with density matrices
$\{|\psi_\alpha\rangle \langle\psi_\alpha|\}_{\alpha=1}^N$ and 
probabilities $\{\lambda_\alpha\}_{\alpha=1}^N$.  They correspond to the
(infinitely) many ways in which the hermitian operator $\hrho$ can be
decomposed as \cite{willi,espagnat,peres,petr}
\BEA
\label{baba}
\hrho=\sum_{\alpha=1}^N\lambda_\alpha
|\psi_\alpha\rangle\langle\psi_\alpha|,
\quad \lambda_\alpha\geq
0,\quad \sum_{\alpha=1}^N\lambda_\alpha=1, 
\EEA 
where $|\psi_\alpha\rangle$ are some normalized 
| but in general not orthogonal| vectors living in the
same $n$-dimensional Hilbert space ${\cal H}$ \footnote{
Normalization and belonging to ${\cal H}$ are necessary for
$|\psi_\alpha\rangle\langle\psi_\alpha|$ to describe some ensemble of
the systems $\S$.}, and where $|\psi_\alpha\rangle\langle\psi_\alpha|$
are distinct.

The eigenresolution (\ref{resolrho}) is only a particular case of
(\ref{baba}), and if now the ensemble $\E(\hrho)$ was prepared by one
of the ways corresponding to (\ref{baba}) with non-orthogonal
$|\psi_\alpha\rangle$, the constituents of $\E(\hrho)$ come from the
subensembles $\E(|\psi_\alpha\rangle \langle\psi_\alpha|)\}$ and the
observable $\hat{B}$ has in general no any definite value for these
subensembles.

The above discussion allows to conclude with two related features of 
a homogeneous ensemble: {\it i)} a single member of such an ensemble does
not by itself define a subensemble;  {\it ii)} the ensemble cannot be 
thought to consist of definite subensembles. 

\subsubsection{Pure-state ensembles.}

The description of a homogeneous ensemble via pure density matrices,
$\hrho^2=\hrho$,
has several special features.  

First of all, it is seen from (\ref{baba}) that for a pure state
$\hrho=|\psi\rangle\langle\psi|$ in the RHS of representation
(\ref{baba}) only one term shows up: $|\psi\rangle\langle\psi|=
|\psi\rangle\langle\psi|$ 
\footnote{\label{a}
This can also be deduced from a more general result: for any
$|\psi_\alpha\rangle$ that can appear in (\ref{baba}), either
$\hrho=|\psi_\alpha\rangle\langle\psi_\alpha|$, or
$|\psi_\alpha\rangle$ is orthogonal to the linear space formed by the
eigenvectors of $\hrho$ corresponding to eigenvalue zero. Indeed, let
$|0\rangle$ be one such eigenvector, then $\langle
0|\hrho|0\rangle=\sum_\alpha\lambda_\alpha\,|\langle
0|\psi_\alpha\rangle|^2=0$; thus $|\langle 0|\psi_\alpha\rangle=0$ for
$\lambda_\alpha>0$.}.
Thus, pure-state ensembles cannot be
prepared via mixing of other ensembles of the system $\S$, or, put
differently, pure-state ensembles are irreducible.

Second, this description is the maximally {\it complete} one possible
in quantum mechanics.  This known thesis can be substantiated as
follows. First one notes from (\ref{dhol}, \ref{duduk}) that for a
fixed $\hrho$ dispersionless observables form a linear space: if two
operators are dispersionless, so is their sum, and multiplication by a
number conserves the dispersionless feature.

From (\ref{duduk}) and Appendix \ref{f7} one sees that if the
mixed density matrix $\hrho$ has $k$, $1\leq k\leq n$, non-zero
eigenvalues ($n$ being the dimension of the Hilbert space ${\cal H}$),
then the dimension of the linear space formed by the corresponding
dispersionless observables is equal to
\BEA N_k=
(n-k)^2+1.
\EEA
This number is maximal for $k=1$, that is, for pure density matrices.
In other words, pure density matrices provide definite values for a
larger set of observables than mixed density matrices
\footnote{For $k=n$ we get $N_k=1$, since in this case only operators
proportional to unity are dispersionless. For $n=2$ and $k=1$,
$N_k=2$: all dispersionless observables for a two-dimensional pure
density matrix $|\psi\rangle\langle\psi|$ can be represented
as $\alpha |\psi\rangle\langle\psi|+\beta |\psi_\bot\rangle\langle\psi_\bot|$,
where $\langle\psi|\psi_\bot\rangle=0$, and where $\alpha$ and $\beta$
are two independent real numbers. }. 
For a mixed state all dispersionless observables have to be degenerate.

Though the above two features of irreducibility and completeness
create a conceptual difference between pure and mixed density
matrices, this should certainly not be taken as an invitation to prescribe pure
density matrices to a single system, reserving the mixed ones for
ensembles; further reasons for this are analyzed in
Refs.~\cite{park,espagnat,peres,willi,petr,abnswedish}~
\footnote{Among reasons we find convincing is the analysis
of the quantum measurement process \cite{abnswedish}.}.

\subsubsection{Inhomogeneous ensembles.}

A mixed density matrix $\hrho$ can also describe inhomogeneous ensembles.  Such
an ensemble $\E_{\rm i}$ is a collection of homogeneous subensembles
$\{\,\E(\hrho_\alpha)\,\}_{\alpha=1}^N$ with probabilities
$\{\,\lambda_\alpha\,\}_{\alpha=1}^N$, so that each single system from
$\E_{\rm i}$ is known to be taken from the ensemble $\E(\hrho_\alpha)$
with probability $\lambda_\alpha$, $\alpha=1,..,N$.  Obvious cases are when
the subensembles $\E(\hrho_\alpha)$ are separated in space or in time,
or by means of some other classical quantity.

Inhomogeneous ensembles are typically prepared by means of selective
measurements \footnote{These measurements need not be done on the systems
$\S$ directly, they can be indirect as well. Imagine an ensemble of two
spin-$\frac{1}{2}$ particles described by pure density matrix
$|\psi\rangle \langle\psi|$, where
$|\psi\rangle=\frac{1}{\sqrt{2}}(|+\rangle_{1}\otimes|+\rangle_{2}
\,+\,|-\rangle_{1}\otimes|-\rangle_{2})$, and where $|\pm\rangle_{1,2}$
are the eigenvectors of $\hat{\sigma}_z^{(1,2)}$ with 
eigenvalues $\pm 1$ for the first and
second particle, respectively.  One can now measure
$\hat{\sigma}_z^{(1)}$, and keep both the results of these measurements and
the order of their appearance (thus, one keeps a sequence of random
numbers $\pm 1$).  For the subensemble of the second spin this amounts
to preparation of inhomogeneous ensemble $\{ \frac{1}{2},
|+\rangle_{2}\,_2\langle +| ~;~ \frac{1}{2}, |-\rangle_{2}\,_2\langle -|
\}$. }. 
In that case the above classical quantity is the
corresponding record of the macroscopic apparatus by which this
measurement was done.  Below in section \ref{POVM} we describe in
detail how an initially homogeneous ensemble can be separated into
subensembles by means of a measurement.

The inhomogeneous ensemble $\E_{\rm i}$ is still described by the
overall density matrix $\hrho=\sum_{\alpha=1}^N\lambda_\alpha
\hrho_\alpha$, but in contrast to the homogeneous situation this is not
the full description. The latter is provided by the list \BEA
\{\,\lambda_\alpha,\, \hrho_\alpha\,\}_{\alpha=1}^N.  \EEA So more
information is known about the inhomogeneous ensemble $\E_{\rm i}$ then
only $\hrho$. 
If the inhomogeneous ensemble is just a combination of homogeneous ones,
this is obvious.
If the inhomogeneous ensemble was prepared by means of 
a measurement, then the above information results from the measurement
carried out and from selection of the outcomes (see more details in
section \ref{tarzan} below).

\subsubsection{Prescribed ensemble fallacy.} 

This fallacy rests on forgetting the difference between homogeneous and
inhomogeneous ensembles \cite{kok,peres}. In spite of explicit warnings
\cite{landau}, the fallacy frequently appears in applications and
interpretations of quantum statistical physics.  Consider, for example,
the basic tool of statistical physics, the equilibrium ensemble
described by the Gibbsian density matrix (\ref{gibbs}). It is typically
obtained by thermalization process, that is, due to interaction with a
thermal bath.  One sometimes hears with respect to this ensemble that it
represents the system being in states of definite energy with the
corresponding probabilities $p_k$. This is a valid description of the
ensemble only after the measurement of energy $\hH$ has been done,
something which is by itself not typical in applications. Moreover, as
we recalled above and below, one can choose to make a different
measurement, and then the interpretation in terms of definite energies
will be explicitly wrong. The reason of why some applications |though
starting from the above incorrect premise| do not lead to contradictions
is clear: they use this premise merely for ``explanations of what
actually happens'', while in real calculations and comparisons with
experiment only the density matrix (\ref{gibbs}) is employed.

\subsection{Fluctuations of work.}
\label{qqff}

Once the properties of quantum ensembles are clarified, we can proceed
with the quantum definition of fluctuations of work.  The most reasonable way
to define this concept in the quantum situation is to proceed along
the same lines as in classics, taking into account when needed the
differences between quantum and classical ensembles.

It is convenient to separate the definition into the following steps.

{\it 1.}  The initial ensemble $\E(\hrho)$ is homogeneous, since it
was prepared by means of a thermal bath.  With help of a suitable
measurement (see section \ref{POVM} for details), one separates
$\E(\hrho)$ into irreducible, maximally complete subensembles
$\{\,\E(|\psi_\alpha\rangle\langle\psi_\alpha|)\,\}_{\alpha=1}^N$ with
probabilities $\{\,\lambda_\alpha\,\}_{\alpha=1}^N$, so that the
resulting inhomogeneous ensemble is still described by
the same density matrix $\hrho$ and thus (\ref{baba}) is valid.  

In the quantum situation irreducible, maximally complete subensembles
are described by pure density matrices $|\psi\rangle\langle \psi|$, as
we recalled above. The important point is that
these subensembles play here the same role as the single systems
for the classical definition of fluctuations of work.

Note that once it is understood that the initial ensemble $\E(\hrho)$
is homogeneous and that measurements are anyhow needed to make it
inhomogeneous, we have to admit any measurement which will produce
pure-state ensembles, even those with non-orthogonal
$|\psi_\alpha\rangle$'s.

Recall that the present step of preparing an inhomogeneous ensemble
out of the initial homogeneous one is absent in the classical
situation, simply because there are no essentially inhomogeneous
classical ensembles.

{\it 2.} This step almost literally repeats its classical analog.  The
single systems from each subensemble
$\E(|\psi_\alpha\rangle\langle\psi_\alpha|)$ interacts with the work
source which realizes the same thermally isolated process on each
single system from each subensemble.

The evolution of the corresponding subensemble
during the cyclic process between times $0$ and $\tau$
is given by the von Neumann equation
\BEA 
i\hbar\frac{\d}{\d t}{\hrho_\alpha}(t)=[\,\hH(t),\hrho_\alpha(t)\,],
\qquad \hrho_\alpha(0)=|\psi_\alpha\rangle\langle\psi_\alpha|\\
\hrho_\alpha(\tau)=\hU_\tau\,\hrho_\alpha(0)\,\hU_\tau^\dagger.
\EEA 

{\it 3.} In analogy with the corresponding classical step we define the
work $w_\alpha$ done on the subensemble $\alpha$ via Eq.~(\ref{work0}):
\BEA
w_\alpha&&={\rm tr}\left(\,
\Del\,|\psi_\alpha\rangle\langle\psi_\alpha|\,\right)\nonumber\\
&&=\langle\psi_\alpha(\tau)|\hH|\psi_\alpha(\tau)\rangle
-\langle\psi_\alpha(0)|\hH|\psi_\alpha(0)\rangle.
\label{qf}
\EEA 
This is the average energy decrease of the mechanical degree of
freedom of the work source due its interaction with the corresponding
subensemble.  Thus $w_\alpha$ has the meaning of work by itself, but
it is a quantity that had to be averaged over the subensemble.  The
probability of $w_\alpha$ is equal to $\lambda_\alpha$, since, as seen
from (\ref{baba}), this is the probability by which the corresponding
pure subensemble enters the overall ensemble described by
$\hrho$.

Thus we defined a random c-number quantity work $w$ with realizations
$w_\alpha$ and probabilities $\lambda_\alpha$:
\BEA
\label{suliko}
w=\left\{w_\alpha\,,\,\lambda_\alpha\right\}_{\alpha=1}^N.
\EEA

As follows from (\ref{quantwork}, \ref{baba})
the work done on the overall ensemble is equal to the weighted average
over the pure subensembles:
\BEA
W=\sum_{\alpha=1}^N\lambda_\alpha w_\alpha.  
\label{gi1}
\EEA

Eq.~(\ref{gi1}) remains true for any initial ensemble. It is
straightforward to see that our definition of fluctuations of work can
be applied to any initial ensemble and not only to that described by
the Gibbsian density matrix (\ref{gibbs}).

The thus defined fluctuations of work do depend on the pure ensembles
$\{|\psi_\alpha\rangle\langle\psi_\alpha|\}_{\alpha=1}^N$, defined
uniquely once the measurement separating the overall ensemble into pure
subsensembles is specified. Strictly speaking, what we defined as
fluctuations are the ones between subensembles (inter-subensemble
fluctuations). Within the standard quantum theory we do not know how to
define fluctuations of work inside of a irreducible subensemble. There
were in literature some attempts in this direction, which are described
in section \ref{others}. However, they do not satisfy the natural
conditions on fluctuations of work, as outlined in the Introduction
(arbitrary initial state; proper physical meaning).
In particular, the approach based on the ``operator of work''
is not applicable, since we will explain that this operator does not
satisfy the proper criteria ~
\footnote{Thus if these fluctuations exist, 
and we assume they do, their description seems to be 
outside of today's theories.
It might be of some interest to see whether more detailed definitions of
fluctuations of work can be given in theories of subquantum mechanics,
e.g., Bohmian or Nelsonian mechanics.}. 

If there is no interaction with any work source, that is, the
Hamiltonian $\hH$ is time-independent, and if in addition $[\hrho,\hH]
=0$, then the ensemble described by $\hrho$ is stationary: all
(one-time) averages are time-independent. Now note that the stationary
ensemble can be decomposed into non-stationary subensembles, since in
general $[| \psi_\alpha\rangle\langle\psi_\alpha|\,,\,\hH]\not =0$.
This is clearly impossible for a classical ensemble, but in the
context of fluctuations of work this fact implies nothing pathologic,
since work is defined for any initial ensemble, not only for
stationary ones. It is checked from (\ref{qf}) that if there is no
interaction with the work-sources, then $\Omega\equiv 0$, and all possible
realizations of work are zero.

{\it 4.} Note that for macroscopic systems it is not realistic to
have available measurements producing pure-state subensembles, since
the directly availabale measurements are only those of macroscopic
quantities which are typically degenerate. In this case we may need 
to apply a coarse-grained
definition of fluctuations of work, where the initial mixed ensemble
is separated into mixed subensembles described by density matrices
$\hat{\sigma}_\gamma$ 
($\hat{\sigma}^2_\gamma\not=\hat{\sigma}_\gamma$)
\BEA
\label{cg}
\hrho=\sum_\gamma\nu_\gamma
\hat{\sigma}_\gamma,\quad
\nu_\gamma\geq 0,\quad
\sum_{\gamma}\nu_\gamma=1.
\EEA
The definition then proceeds as above with obvious changes (e.g.,
$|\psi_\alpha\rangle\langle\psi_\alpha|\to \hat{\sigma}_\gamma$ in 
(\ref{qf})~).

This is a coarse-grained definition, since the realizations of work
${\rm tr}\,(\Del\,\hat{\sigma}_\gamma)$ can be reduced to more fundamental 
ones, i.e., each of them can be presented as a convex sum of ${\rm
tr}\,(|\psi_\alpha\rangle\langle\psi_\alpha|\,\Del\,)$.
As a consequence fluctuations of work |as quantified, e.g., by dispersion
of work defined and discussed in section \ref{dispersion}| are maximal for
pure-state decompositions (more details on this are found in section
\ref{dispersion1}). 

\subsection{Separation of a homogeneous ensemble into
pure subensembles by filtering outcomes of a POVM measurement.}
\label{POVM}
\label{tarzan}

\subsubsection{Positive Operator Valued Measurements.}

It is now our purpose to discuss how precisely one separates with help
of measurements an initial homogeneous ensemble $\E(\hrho)$ into pure
(necessarily homogeneous) subensembles.

The most general type of a quantum measurement corresponds to
Positive Operator Valued Measure (POVM) \cite{peres,willi} defined
via $N$ operators $\hat{G}_\alpha$ | not necessarily orthogonal |
living in the $n$-dimensional Hilbert space ${\cal H}$ 
and satisfying the completeness relation
\BEA
\label{comp}
\sum_{\alpha=1}^N\hat{G}^\dagger_\alpha\hat{G}_\alpha=
\hat{1}.
\EEA

The most standard measurements of an observable $\hat{A}$ living in
the $n$-dimensional Hilbert space ${\cal H}$ and having non-degenerate
spectrum $\{a_\alpha\}_{\alpha=1}^n$ are included in (\ref{comp}),
since now $N=n$ and $\{\hat{G}_\alpha\}_{\alpha=1}^n
=\{|a_\alpha\rangle\langle a_\alpha|\}_{\alpha=1}^n$, where the latter
is the set of orthonormal eigenvectors of $\hat{A}$.  If the spectrum
of $\hat{A}$ happens to have degeneracies, so that each eigenvalue
$a_\alpha$ has multiplicity $n_\alpha$, then $\hat{G}_\alpha$ is the
$n_\alpha$-dimensional projector on the subspace formed by
$n_\alpha$ linearly independent eigenvectors of $\hat{A}$ which
correspond to the eigenvalue $a_\alpha$. Here $N\leq n$ is equal to
the number of distinct eigenvalues of $\hat{A}$.

If the measurement described by (\ref{comp}) is done on the ensemble
described by a density matrix $\hrho$, then the result $\alpha$ is found
with probability
\BEA
\label{durman}
\lambda_\alpha=\tr(\hat{G}^\dagger_\alpha\hat{G}_\alpha\hrho)=
\tr(\hat{G}_\alpha\hrho\hat{G}^\dagger_\alpha),
\EEA
where $\lambda_\alpha\geq 0$ and $\sum_{\alpha=1}^N\lambda_\alpha=1$,
due to (\ref{comp}).
After selecting results of the measurements
referring to the outcome $\alpha$ 
one has the (sub)ensemble of systems described by a density matrix
\BEA
\label{yale}
\hrho'_\alpha=\frac{\hat{G}_\alpha\hrho\hat{G}^\dagger_\alpha}
{\tr(\hat{G}^\dagger_\alpha\hat{G}_\alpha\hrho)}.
\EEA
This subensemble occura with probability $\lambda_\alpha$ as given by (\ref{durman}),
simply because this is the probability of the outcome $\alpha$.
The overall post-measurement inhomogeneous
ensemble thus
{\it consists of} $N$ subensembles each of which has a density matrix
(\ref{yale}) and probability (\ref{durman}). 
The density matrix of the
overall post-measurement ensemble is 
\BEA
\hrho'=\sum_{\alpha=1}^N\lambda_\alpha\hrho'_\alpha.
\EEA

POVM's are related to more usual projective measurements, where
$\hat{G}_\alpha$ are mutually orthogonal,
$\hat{G}_\alpha\hat{G}_\beta=\hat{G}_\alpha\delta_{\alpha\beta}$,
projections into eigenspaces of some hermitian operator: according to
Neumark's theorem \cite{peres,willi} every POVM can be realized as
some (non-unique) projective measurement in a larger Hilbert space,
that is involving additional degrees of freedom.  A detailed
discussion of this theorem and various versions 
is given in \cite{peres,willi}.  In Appendix \ref{povm} we shall
discuss an example of it that is relevant for our purposes.  

\subsubsection{}
Applying a POVM measurement, one now wishes to separate the mixed
quantum ensemble described by the density
matrix $\hrho$ into pure subensembles. The density
matrix $\hrho'$ of the overall post-measurement ensembles 
should then coincide with
$\hrho$ given in (\ref{resolrho}) or (\ref{gibbs}), while $\hrho'_\alpha$
appearing in (\ref{yale}) should be pure:
\BEA
\hrho'_\alpha=|\psi_\alpha\rangle\langle\psi_\alpha|.
\EEA
Then the density matrix (\ref{gibbs}) is decomposed as in
(\ref{baba}).

Let us first see which $\{\lambda_\alpha\}_{\alpha=1}^N$ and
$\{|\psi_\alpha\rangle\langle\psi_\alpha|\}_{\alpha=1}^N$ are allowed
to enter in (\ref{baba}), and then we shall discuss which specific
measurements should be done to achieve the actual separation.

It will prove useful to write (\ref{baba}) in an equivalent way
\BEA
\hrho=\sum_{\alpha=1}^N
\widetilde{ |\psi_\alpha\rangle } \widetilde{ 
\langle\psi_\alpha|},\quad \widetilde{|\psi_\alpha\rangle }
\equiv\sqrt{\lambda_\alpha}|\psi_\alpha\rangle,
\label{dodo}
\EEA 
since it will allow us to focus on $\widetilde{ |\psi_\alpha\rangle }$,
keeping in mind that the probabilities $\lambda_\alpha$ can always
be recovered via $\lambda_\alpha=\widetilde{ \langle\psi_\alpha }
|\widetilde{\psi_\alpha\rangle}$.

According to the ensemble classification theorem \cite{erwi,jaynes,wotti,gigi},
if one has
\BEA
\label{kara}
\widetilde{ 
|\psi_\alpha\rangle}=\sum_{k=1}^n
M_{\alpha k}\,\sqrt{p_k}|p_k\rangle,
\EEA
where $\{p_k\}_{k=1}^n$, $\{|p_k\rangle\}_{k=1}^n$ are the eigenvalues
and eigenfunctions of the density matrix $\hrho$, and where $M_{\alpha
k}$ are complex numbers satisfying
\BEA
\label{kazan}
\sum_{\alpha=1}^N M_{\alpha k}M^*_{\alpha j}=\delta_{ij},
\quad k,j=1,...,n,
\EEA
then Eq.~(\ref{dodo}) becomes $\hrho=\sum_{k=1}^n
p_k|p_k\rangle\langle p_k|$, as it should
\footnote{\label{vakho} Note that any vector
$\widetilde{|\psi_\alpha\rangle}$ having
$\widetilde{\langle\psi_\alpha } |\widetilde{ \psi_\alpha\rangle}<1$
and living in the Hilbert space formed by the eigenvectors
of $\hrho$ corresponding to its non-zero eigenvalues, 
can appear in at least one separation (\ref{dodo}) of
$\hrho$. This follows from (\ref{kara}).}.

The converse appears to be true as
well: any decomposition (\ref{dodo}) admits a
representation (\ref{kara}) with some complex numbers $M_{\alpha k}$
satisfying (\ref{kazan}) 
\footnote{\label{bulo} To prove this part of the statement, 
recall Footnote \ref{a},
expand $
\widetilde{ |\psi_\alpha\rangle} $ over the eigenbase
$|p_k\rangle$ of $\hrho$: $\widetilde{|\psi_\alpha\rangle
}=\sum_{k=1}^n\langle p_k \widetilde{|\psi_\alpha\rangle
}\,|p_k\rangle$, substitute this into (\ref{dodo}), and then deduce
(\ref{kazan}) using the orthonormality and completness of the above
base in the Hilbert space ${\cal H}$: $\sum_{\alpha=1}^N\langle p_k
\widetilde{|\psi_\alpha\rangle }\, \widetilde{\langle\psi_\alpha|
}p_l\rangle=\langle p_k|\hrho|p_l\rangle =\delta_{kl}\,p_k$.  Thus,
any decomposition (\ref{dodo}, \ref{baba}) can be constructed via
(\ref{kara}) and
$M_{\alpha k}\langle p_k
\widetilde{|\psi_\alpha\rangle }/\sqrt{p_k}$
satisfying (\ref{kara}). 
If some eigenvalues of $\hrho$ are equal to zero, than 
the above construction should be restricted to eigenvectors of $\hrho$
corresponding to its non-zero eigenvalues.
}.

Note that Eq.~(\ref{kara}) implies the following formula for the
probabilities $\lambda_\alpha$:
\BEA
\label{afgan}
\lambda_\alpha=\widetilde{\langle\psi_\alpha }
|\widetilde{ 
\psi_\alpha\rangle}=
\sum_{k=1}^n|M_{\alpha k}|^2 p_k.
\EEA

As seen from (\ref{kazan}), the very possibility of writing Eq.~(\ref{dodo})
implies
\BEA
N\geq n, 
\label{dolores}
\EEA
since $M_{\alpha k}$ can be
viewed as $n$ different $N$-component orthonormal vectors. The rectangular
matrix $\{\{M_{\alpha k}\}_{\alpha=1}^N\}_{k=1}^{n}$ can be completed
to a unitary $N\times N$ matrix by adding suitable elements.

It is now straightforward to see which POVM 
can be taken to achieve the decomposition (\ref{dodo}). Take, for example,
\BEA 
\label{dandalosh}
G_{\alpha}=
\frac{|\widetilde{\psi_\alpha\rangle}
\widetilde{\langle\psi_\alpha|}\,\hrho^{-1/2} }
{\sqrt{\langle\widetilde{\psi_\alpha}\,|\,
\widetilde{\psi_\alpha\rangle}}}=
\sqrt{\lambda_\alpha}\,|\psi_\alpha\rangle\,\langle\psi_\alpha|\,
\hrho^{-1/2},
\EEA
where $|\psi_\alpha\rangle$ is defined in (\ref{dodo}).
Note that the converse appears to be true as well. For given POVM
(\ref{comp}) with
\BEA
\label{chibukh1}
\hat{G}^\dagger_\alpha\hat{G}_\alpha=|\pi_\alpha\rangle\langle
\pi_\alpha|,
\EEA
where $|\pi_\alpha\rangle$ satisfying
\BEA
\label{sasan}
\hat{1}=\sum_{\alpha=1}^N|\pi_\alpha\rangle\langle\pi_\alpha|,
\EEA
have to be neither orthogonal, nor normalized 
\footnote{\label{f99}If one assumes in Eq.~(\ref{sasan}) that $|\pi_\alpha\rangle$ are
  normalized, $\langle\pi_\alpha|\pi_\alpha\rangle=1$, then this leads
  to orthogonality:
  $\langle\pi_\beta|\pi_\alpha\rangle=\delta_{\alpha\beta}$. Indeed,
  denoting $\hPi_\alpha=|\pi_\alpha\rangle\langle\pi_\alpha|$, one gets
  $\sum_{\alpha\not=\beta}^N\left(\hPi_\alpha\hPi_\beta
  \right)^\dagger\,\left(\hPi_\alpha\hPi_\beta
  \right)=\sum_{\alpha\not=\beta}^N\hPi_\beta\hPi_\alpha\hPi_\beta=
  \hPi_\beta(1-\hPi_\beta)\hPi_\beta=0$. Since
  $\left(\hPi_\alpha\hPi_\beta
  \right)^\dagger\,\left(\hPi_\alpha\hPi_\beta \right)$ is non-negative
  by construction, one concludes $\hPi_\alpha\hPi_\beta=0$ for
  $\alpha\not =\beta$.}, one can construct a representation 
(\ref{dodo}, \ref{baba}) of $\hrho$ as
\BEA
\hrho=\sum_{\alpha=1}^N\hrho^{1/2}\,
|\pi_\alpha\rangle\langle\pi_\alpha|\,\hrho^{1/2}.
\label{lotos}
\EEA

Thus, we have seen how all possible decompositions of a mixed
ensemble into pure subensenbles can be constructed via suitable
measurements. 

We stress that the decompositions into a specific set of subensembles
is related to a physical measurement, rather than to a mathematical choice. 

\subsubsection{Preparation versus measurements.}

To avoid possible confusions we recall once again that the above
separation procedure corresponds to {\it preparation} of the
inhomogeneous ensemble 
$\{\,\lambda_\alpha\,,\,
\E(|\psi_\alpha\rangle\langle\psi_\alpha|)\,\}_{\alpha=1}^N$ 
with $\hrho=\sum_\alpha\lambda_\alpha
|\psi_\alpha\rangle\langle\psi_\alpha|$, starting from the initial
homogeneous ensemble $\E(\hrho)$.
Though this preparation was based on a suitable measurement process,
we were not interested by some aspects usually associated with it.
For example, we did not keep track of the pointer
variable of the measuring apparatus, which obviously should be the
main goal of any measurement process studied for its own purposes
\cite{abnswedish}.
We were more interested by the influence of the measurement process
on the final state of the system $\S$, which is the basic
characteristic feature of the preparation process in quantum mechanics
\cite{willi}.

\subsection{Discussion.}

There are several questions on the physical meaning of the proposed
definition of fluctuations of work that we decided to discuss 
separately.

{\it Question 1.}  Among all decompositions (\ref{baba}) 
of the Gibbsian density matrix $\hrho$, there is a
unique one (up to accidental degeneracies of the spectrum) given by
the eigenvectors of $\hrho$ and realized via measurement of the
Hamiltonian $\hat{H}$.  Then the energy has a definite value on each
subensemble.  Should not one therefore restrict the definition of
fluctuations of work to this separation only?

{\it Answer 1.} There are at least two reasons why the answer is no.
First, even if the energy has a definite value initially, it will in
general not have any definite value at the final moment, since an
eigenvector of the initial Hamiltonian may evolve into a superposition
of eigenvectors.  Thus, there are no special reasons to insist on the
feature of separating with respect to energy.  Second, more general
separations are anyhow necessary to define fluctuations of work for an
arbitrary ensemble, which cannot be decomposed into subensembles with
each of them having a definite value of energy.

{\it Question 2.}  Is the orthogonal separation not special by the
fact that various ensembles are described by orthogonal pure density
matrices, and can thus be discriminated unambigiously?  

{\it Answer 2.} By definition any POVM is connected with an unambigious
descrimination of its different outcomes. This can be additionally
clarified by looking at the example of the projective realization of a
POVM presented in Appendix \ref{povm}, where various subensembles
constructed after the measurement are seen to be described by orthogonal
wave-functions in the composite Hilbert space ${\cal H}\otimes {\cal
H}'$.  The above question mixes the present situation with a different
one, where one is given a single system coming from one of two ensembles
having non-orthogonal density matrices, and is requested to determine by
means of a measurement from which ensemble it is coming.  Then, indeed,
no measurements can ensure unambigious discrimination \cite{peres}.

{\it Question 3.}  The authors prescribe to the viewpoint that even
pure density matrices (wave functions) describe an ensemble of quantum
systems and not a single system, as some people like to think. How the
proposed definition will change, if one would wish to insist on the
latter interpretation of quantum mechanics?

{\it Answer 3.}  The necessity of prescribing even the pure density
matrices to ensembles of quantum systems was stressed in
\cite{espagnat,park,peres,willi}. In particular, it is needed for the
consistent solution of the quantum measurement problem \cite{abnswedish,willi}.
It is also known that with respect to certain
aspects of quantum theory the prescription of pure density matrices to
a single system is relatively harmless. We do not have space to
discuss in detail what are those aspects and what precisely is meant
by ``relatively harmless''.  At any rate, we do not advise to make the
latter prescription, and the readers who wish to do that have to 
proceed on their own risk. We may mention that the definition 
of fluctuations of work
remains then basically unchanged, but becomes conceptually closer
to its classical analog, since now in defining fluctuations of work
one assumes to operate with single systems both in quantum and classical
situations.

\section{Dispersion of work.}
\label{dispersion}

The most direct quantity that characterizes how the 
realizations (\ref{qf}) of the random quantity work 
are spread around their mean
$W=\sum_{\alpha=1}^N\lambda_\alpha w_\alpha$, is the 
(inter-subensemble) dispersion
\BEA
\label{wd}
\del &=& \sum_{\alpha=1}^N\lambda_\alpha 
\left[\,\langle\psi_\alpha|\Del|\psi_\alpha\rangle
-\tr (\Del \hrho)\,\right]^2\\
&=&\sum_{\alpha=1}^N\lambda_\alpha (w_\alpha-W)^2=
\sum_{\alpha=1}^N\lambda_\alpha w_\alpha^2-W^2.
\label{workdispersion}
\EEA
In contrast to $W$, this quantity depends explicitly on the subensembles
used to define $w_\alpha$ in (\ref{qf}). So it depends explicitly on the
physical process that separated the initial ensemble into subensembles.

It is useful to
to determine the maximal $\del_{\rm max}$ and the minimal
$\del_{\rm min}$ values of $\del$ over all possible decompositions
$\{|\psi_\alpha\rangle\langle\psi_\alpha|,\,\lambda_\alpha\}_{\alpha=1}^N$
corresponding to the fixed $\hrho=\sum_{\alpha=1}^N\lambda_\alpha
|\psi_\alpha\rangle\langle\psi_\alpha|$.
According to (\ref{lotos}) these extremizations can be equally well
carried out over all possible decompositions of unity in our $n$-dimensional
Hilbert space,
\BEA
\label{chibukh2}
\sum_{\alpha=1}^N\hPi_\alpha=\hat{1},\qquad \hPi_\alpha=|\pi_\alpha\rangle
\langle\pi_\alpha|,
\EEA
where $\{|\pi_\alpha\rangle\}_{\alpha=1}^N$ 
have in general to be neither normalized
nor orthogonal. 

Note that dispersions similar to (\ref{wd}), with $\Omega$
corresponding to some other relevant observable, where introduced and
studied in quantum optics, where separation of an ensemble by means of
(continuous) measurements are well-known and were studied both
experimentally and theoretically; see \cite{petr} for a review. The
results we present below on the minimal and maximal values of the
dispersion $\delta w^2$ do not depend on the details of $\Omega$ and
can thus be useful in general.

\subsection{Maximal dispersion of work.}
\label{noopofwork}
\label{maxi}
\label{dispersion1}

The maximization of $\del$ over all possible separations (\ref{baba},
\ref{chibukh2})
for given $\hrho$ and $\Del$ is carried out in Appendix \ref{barbi}.
The result is 
\BEA
\label{dard1}
\del_{\rm max}=\sum_{i,k=1}^n
\frac{2 p_i p_k}{p_i+p_k}\,|\langle\ep_k|\Del |\ep_i\rangle|^2-W^2\\
=2\,\int_0^\infty\d s\,{\rm tr}\left [
\left(
\Del\,\hrho\,e^{-s\hrho}
\right)^2
\right]-W^2.
\label{dard11}
\EEA
This maximum is reached on $\{|\pi_\alpha\rangle\}_{\alpha=1}^n$ 
being the eigenvectors of a hermitian operator 
\BEA
\label{dard2}
\hat{X}=\sum_{i,k=1}^n
\frac{2 p_i p_k\,\,\langle p_i|\Del |p_k\rangle}{p_i+p_k}\,
| p_i\rangle\langle p_k|,
\EEA
where $p_k$ and $| p_k\rangle$ are the eigenvalues and eigenvectors
of $\hrho$, as defined by (\ref{resolrho}).

Only when $\hrho$ and $\Del$ commute, $[\hrho,\Del]=0$, the maximal dispersion
(\ref{dard1}) reduces to the more usual expression 
${\rm tr}[\hrho\,\Del^2]-[{\rm tr}(\hrho\,\Del)]^2$. This and related questions are
discussed in more detail around Eqs.~(\ref{espanol}, \ref{camaradas}).

The maximal dispersion (\ref{dard1}, \ref{dard11})
provides an upper bound for the dispersion of work defined 
in a coarse-grained way; see the discussion around Eq.~(\ref{cg}).
Indeed according to that discussion the coarse-grained dispersion 
of work defined with respect to separation of ${\cal E}(\hrho)$ 
to mixed-state subensembles reads
\BEA
\delta w_{\rm cg}^2=\sum_\gamma \nu_\gamma ({\rm tr}(\hat{\sigma}_\gamma\Del)-W)^2.
\EEA
Note a decomposition of $\hat{\sigma}_\gamma$ into some set of pure-state subensembles, 
$\hat{\sigma}_\gamma=\sum_\alpha \mu^{(\gamma)}_\alpha |\psi^{(\gamma)}_\alpha
\rangle\langle \psi^{(\gamma)}_\alpha|$, where $\mu^{(\gamma)}_\alpha$
are the corresponding probabilities with
$\sum_\alpha \mu^{(\gamma)}_\alpha=1$. One now finds that the dispersion 
$\delta w^2$ defined
as in Eqs.~(\ref{wd}, \ref{workdispersion}), that is,
via the separation of the ensemble ${\cal E}(\hrho)$ 
into pure-state subensembles
$\hrho=\sum_{\alpha,\gamma} 
\nu_\gamma\mu^{(\gamma)}_\alpha |\psi^{(\gamma)}_\alpha
\rangle\langle \psi^{(\gamma)}_\alpha|$, is always not smaller than 
$\delta w_{\rm cg}^2$:
\BEA
\del-\delta w_{\rm cg}^2&\geq&
\sum_{\alpha,\gamma} 
\nu_\gamma\mu^{(\gamma)}_\alpha \,(\,\langle \psi^{(\gamma)}_\alpha|\Del|
\psi^{(\gamma)}_\alpha\rangle-W\,)^2
-\delta w_{\rm cg}^2\nonumber\\
&&=\sum_{\alpha,\gamma} 
\nu_\gamma\mu^{(\gamma)}_\alpha \,\left(\,\langle \psi^{(\gamma)}_\alpha|\Del|
\psi^{(\gamma)}_\alpha\rangle 
\right.\nonumber\\ &&\left.-
\sum_{\beta} 
\mu^{(\gamma)}_\beta \langle \psi^{(\gamma)}_\beta|\Del|
\psi^{(\gamma)}_\beta\rangle
\right)^2\geq 0.
\EEA

\subsubsection{The behavior of the maximal dispersion
$\del_{\rm max}$ for high and low
temperatures.}

With $\hrho$ given by the Gibbs distribution (\ref{gibbs}, \ref{resolrho},
\ref{pk}), one gets from Eq.~(\ref{dard1})
\BEA
\del_{\rm max}\to 0,\quad {\rm for}\quad
T\to 0,
\EEA
where $T$ is the temperature of the Gibbsian ensemble. This is a
natural result, as for a finite system $\S$ and $T\to 0$ one gets
$\hrho\to |\eps_0\rangle\langle\eps_0|$, where according to
(\ref{resolH}, \ref{resolrho}), $|\eps_0\rangle$ is the common
eigenvector of $\hrho$ and $\hH$ corresponding to the lowest energy
(assuming that the latter is not degenerate). As no separation of a
pure state into subensembles is possible, the work can take only one
value. It is obvious that this is a general feature: the work does not
fluctuate if the initial ensemble is pure. In the same way as in
classics, fluctuations of work are present for mixed ensembles only.
In this respect the dispersion of work is similar to the von Neumann entropy
$S_{\rm vN}=-\tr\hrho\ln\hrho$, which is also equal to zero for pure
density matrices $\hrho$.

For very high temperatures, where $\hrho\simeq \hat{1}/n$, one gets from
(\ref{dard1})
\BEA
\del_{\rm max}=\frac{1}{n}\,\tr\left(
\Del^2 \right).
\label{gamala}
\EEA
It is seen that for high temperatures the maximal dispersion may be 
${\cal O}(1)$, provided that the (positive) eigenvalues of $\Del^2$
are finite and do not scale with $n$.

\subsection{Minimal dispersion of work.}
\label{minimaldispersion}

Here we show that there are decompositions into
subensembles such that for any $\alpha=1,...,N$:
\BEA
w_\alpha=\langle\psi_\alpha|\Del|\psi_\alpha\rangle=
\sum_{\beta=1}^N\lambda_\beta w_\beta=W,
\label{chibukh3}
\EEA
that is, the work does not fluctuate at all.  In particular, this
means that the dispersion $\del$ attains its minimal value equal
to zero.  This fact is contrasting to the classical situation, where
according to points ${\it b}$ and ${\it c}$ in section \ref{messages},
$w(x,p)$ should be negative at least for some values of $(x,p)$, and
the dispersion of work is large at least for sufficiently high
temperatures.

Recall that due to the parametrization (\ref{chibukh1}, \ref{lotos}, \ref{chibukh2}),
Eq.~(\ref{chibukh3}) can be written as
\BEA
\frac{\langle\pi_\alpha|\hrho^{1/2}\,\Del\,\hrho^{1/2}|\pi_\alpha\rangle}
{\langle\pi_\alpha|\hrho|\pi_\alpha\rangle}={\rm tr}(\,\hrho\,\Del\,),
\EEA 
where $\{|\pi_\alpha\rangle\}_{\alpha=1}^N$ with $N\geq n$
have to satisfy (\ref{chibukh2}). This is equivalent to
\BEA
\label{chibukh33}
0&=&\langle\pi_\alpha|\hat{Y}|\pi_\alpha\rangle,\\
\hat{Y}&\equiv&\hrho^{1/2}\,\hPi_\alpha\,\hrho^{1/2}-{\rm tr}\,
[\,\Del\,\hrho\,]\,\hrho, 
\EEA
where $\hat{Y}$ is hermitian and traceless:
\BEA
{\rm tr}\,\hat{Y}=0.
\label{mumu}
\EEA

We now intend to show that in the Hilbert space ${\cal H}$ there are
orthonormal bases $\{|\pi_i\rangle\}_{i=1}^n$ which for the given
$\hat{Y}$ do satisfy to (\ref{chibukh3}, \ref{chibukh33}). 

\subsubsection{
Some concepts from majorization theory.}

To this end, let us recall some concepts from the mathematical theory
of majorization \cite{major1,major2,major3,ando}. 
For two real vectors $x=(x_1\geq...\geq x_n)$
and $y=(y_1\geq...\geq y_n)$, with their components
arranged in non-increasing way, $y$ is said to majorize $x$,
\BEA
\label{krab1}
x\prec y,
\EEA
if the following conditions are satisfied
\BEA
\label{krab2}
&&\sum_{i=1}^kx_i\leq \sum_{i=1}^ky_i, \qquad
k=1,..,n-1,\\
&&\sum_{i=1}^nx_i=\sum_{i=1}^ny_i.
\label{krab3}
\EEA
Due to Horn's theorem \cite{major1,major2,major3,ando}, 
Eq.~(\ref{krab1}) implies the existence of
a $n\times n$ unitary matrix $\q_{ij}$ such that 
\BEA
x_i=\sum_{k=1}^ny_j\,|\q_{ij}|^2.
\label{horn}
\EEA
The proof of this statement is recalled in Appendix \ref{hornappendix}. 
This proof is constructive, since it allows to gets $\q_{ij}$ starting from
given $x$ and $y$. 

\subsubsection{The minimal dispersion of work is zero.}

Now denote by $(y_1\geq...\geq y_n)$ the eigenvalues of the hermitian
matrix $\hat{Y}$ arranged in non-increasing way. Denote by 
$\{|y_i\rangle\}_{i=1}^n$
the corresponding eigenvectors.
As follows from
(\ref{mumu}, \ref{krab2}, \ref{krab3})
\BEA
(y_1,...,y_n)\,\succ (0,...,0).
\EEA
According to (\ref{horn}) there exists a unitary operator
$\hat{Q}$ in the Hilbert space ${\cal H}$ such that
\BEA
\label{gerasim}
0=\sum_{j=1}^ny_j\, |\langle y_j | \hat{\q} | y_i\rangle|^2
=\langle y_i |\, \hat{\q}^\dagger\,\hat{Y}\,\hat{\q} \,| y_i\rangle.
\EEA
By denoting
\BEA
\hat{\q}\,|y_i\rangle=|\pi_i\rangle,\qquad 
i=1,...,n,
\EEA
we see that (\ref{gerasim}) and the desired statement
(\ref{chibukh33}) are equivalent.

\subsection{Dispersion of work averaged over all separations of 
the ensemble.}

We have obtained the maximal and the minimal values of the dispersion
of work $\del$. It is useful to have a third characteristic
value of $\del$, the dispersion of work for a randomly
chosen separation of the initial ensemble described by $\hrho$ into
pure subensembles. Such a quantity will not depend explicitly on the
measurement used for separation, and thus will help to understand how
typical are the maximal and the minimal values of $\del$.  

Note from Eqs.~(\ref{kara}, \ref{kazan}) that for a given separation of $\hrho$,
that is, for a given representation (\ref{baba}), the pure density
matrices $|\psi_\alpha\rangle\langle\psi_\alpha|$ are 
expressed via elements $M_{\alpha i}$ of a $N\times N$ unitary matrix
$M$ (see the remark after (\ref{dolores})).
We shall define the average dispersion $\del_{\rm av}$ by
assuming that $M$ is random, and then integrating $\del\{M_{\alpha
i}\}$ over all possible unitary $N\times N$ matrices. Since there are
no reasons for introducing {\it a priori} biases, we shall assume for
the above integration the most uniform, unitary-invariant measure
(Haar's measure):
\BEA
\del_{\rm av}=\frac{
\int \prod_{i,\alpha=1}^N\,\d \Re M_{\alpha i}
\,\d \Im M_{\alpha i}\,\Theta\{M_{\alpha i}\}\,\del\{M_{\alpha i}\}
}
{
\int \prod_{i,\alpha=1}^N\,\d \Re M_{\alpha i}
\,\d \Im M_{\alpha i}\,\Theta\{M_{\alpha i}\}
},\nonumber\\
\EEA
where $\Theta\{M_{\alpha i}\}$ comes due to the unitarity constraint
\BEA
\Theta\{M_{\alpha i}\}=
\prod_{\alpha=1}^N\delta \left[
\sum_{i=1}^N|M_{\alpha i}|^2-1\right]
\prod_{\alpha<\beta}^N
\delta \left[
\sum_{i=1}^NM_{\alpha i}M^*_{\beta i}\right].\nonumber
\label{haarmeasure}
\EEA
The rows (or, equivalently, the columns) 
of the matrix $M$ are thus assumed to be
a set of $N$ orthonormalized, uniformly random vectors. The
quantity $\del_{\rm av}$ is calculated in Appendix \ref{haar}:
\BEA
\del_{\rm av}=\int_0^\infty\d s\,\left[\,\prod_{k=1}^n
\frac{1}{1+sp_m}\,\right]\,\left[
\sum_{i=1}^n\left(
\frac{p_i\,\langle\ep_i|\Del|\ep_i\rangle}{1+sp_i}
\right)^2
\right.\nonumber\\
+\left.
\left(\sum_{i=1}^n
\frac{p_i\,\langle\ep_i|\Del|\ep_i\rangle}{1+sp_i}\right)^2\,
\right]-W^2.~~~~
\label{avo}
\EEA
Note that  $\del_{\rm av}$ depends neither on $N$, nor on the 
off-diagonal elements $\langle\ep_i|\Del|\ep_j\rangle$ of $\Del$.

For $\hrho$ having the Gibbsian form (\ref{gibbs}, \ref{resolrho},
\ref{pk}), $\del_{\rm av}$ has the following features for low and
high temperatures $T$. It goes to zero for $T\to 0$ for the same
reasons as $\del_{\rm max}$ does.  In contrast, for very high
temperatures, where $\hrho\simeq 1/n$, one has from (\ref{avo})
\BEA
\del_{\rm av}=\frac{1}{n(n+1)}
\sum_{i=1}^n\langle\ep_i|\Del |\ep_i\rangle^2.
\label{ets}
\EEA
Under the same natural condition that we adopted for studying the
high-temperature behavior of $\del_{\rm max}$, that is, 
$\langle\ep_i|\Del |\ep_i\rangle$
are finite and do not scale with $n$, we see that 
$\del_{\rm av}\propto 1/n$ for $n\gg 1$, which is a typical behavior
for dispersions of fluctuating macroscopic quantities in statistical
physics ~\cite{landau}. Note the difference with the high-temperature
behavior of the maximal dispersion given by Eq.~(\ref{gamala}).

\subsection{The maximal and the average dispersion of work illustrated for 
a two-level system.}
\label{2l}

Let us give concrete expressions of $\del_{\rm max}$ and $\del_{\rm
av}$ for a two-level system $\S$. The initial Gibbsian density matrix
is now a $2\times 2$ diagonal matrix with eigenvalues $p_1$ and
$p_2\leq p_1$ as given by (\ref{resolrho}). The most general matrix
form of the traceless and hermitian operator $\Del$ in this
two-dimensional situation is \BEA
\label{dada}
\Del=
\left(\begin{array}{rr}
\omega & \chi \\
\\
\chi^* & -\omega\\
\end{array}\right).
\EEA
Eqs.~(\ref{dard1}, \ref{avo}) produce then the following expressions
for $\del_{\rm max}$ and $\del_{\rm av}$, respectively:
\BEA
\label{shun}
\del_{\rm max}=\omega^2(1-x^2)\left(
1+\frac{|\chi|^2}{\omega^2}
\right),~~~~\\
\del_{\rm av}=\omega^2(1-x^2)\left[
1-\frac{1}{x^2}\left(1+
\frac{1-x^2}{2x}\,\ln\frac{1-x}{1+x}
\right)
\right],\nonumber\\
\label{katu}
\EEA
where
\BEA
x\equiv p_1-p_2\geq 0,\qquad 1\geq x\geq 0,
\EEA
is a monotonically decreasing function of temperature, as follows from
Eq.~(\ref{pk}). As seen from (\ref{shun}, \ref{katu}), both 
$\del_{\rm max}$ and $\del_{\rm av}$ are monotonically increasing
functions of temperature $T$. It is obvious that $\del_{\rm
max}>\del_{\rm av}$, except for the zero temperature situation
$x=1$, where they are both equal to zero. For very high temperatures,
that is, for $x\to 0$, $\del_{\rm av}=1/3$ in agreement with
(\ref{ets}). Note that off-diagonal elements of $\Del$ increase
$\del_{\rm max}$, while $\del_{\rm av}$ does not depend on
them at all.

\section{There is no direct analog of the classical BK equality
in the quantum situation.}
\label{no}

The discussion in section \ref{minimaldispersion} provides a definite
evidence to think that in contrast to the classical case, the
fluctuations of work in the quantum situation are not controlled by
any {\it direct} analog of the classic BK equality (\ref{gort}). In
the present section we give another illustration of this fact.

Assume for concreteness that the Gibbsian density matrix $\hrho$ in
(\ref{gibbs}) was separated into pure subensembles by means of the
measurement of $\hH$, that is, the subensembles are described by pure
density matrices $\{|\ep_l\rangle\langle\ep_l|\}_{l=1}^n$, where
$\{|\ep_l\rangle\}_{l=1}^n$ are eigenvectors of $\hrho$. 

According to (\ref{qf}) one has for 
realizations of the random quantity work
\BEA
w_l&=&\langle\ep_l|\hU^\dagger_\tau\hH\hU_\tau|\ep_l\rangle-
\ep_l,\\
&=&\sum_{k=1}^nC_{kl}\ep_k-\ep_l,
\quad l=1,...,n,
\EEA
where
\BEA
C_{kl}=
|\langle\eps_k|\hU_\tau|\eps_l\rangle|^2,
\label{dudu1}
\EEA 
is a double-stochastic matrix: 
\BEA
\label{ds}
\sum_{k=1}^nC_{kl}=\sum_{l=1}^nC_{kl}=1.
\EEA
Each of realizations $w_l$ has probability $p_l$, as given by (\ref{pk}).

One now constructs
\BEA
\langle e^{-\beta w}\rangle\equiv
\sum_{l=1}^np_l e^{-\beta w_l}=\frac{1}{Z}
\sum_{l=1}^n e^{-\beta\sum_{k=1}^nC_{kl}\ep_k },
\label{shuka}
\EEA
that is, averages $e^{-\beta w}$ directly as was done in the classical
situation. As shown in Appendix \ref{secul},
\BEA
1-\frac{\beta^2\Delta}{2Z}\,e^{-\beta \ep_{\rm min}}\,
\leq
\langle e^{-\beta w}\rangle
\leq
1-\frac{\beta^2\Delta}{2Z}\,e^{-\beta \ep_{\rm max}},
\label{bazar}
\EEA
\BEA
\Delta&\equiv&
\ep^{\rm T}(1-CC^{\rm T})\ep\nonumber\\
&=&\sum_{k=1}^n\left[\langle\ep_k|\hH|\ep_k\rangle^2-
\langle \ep_k | \hU^\dagger_\tau\,\hH\,\hU_\tau|\ep_k\rangle^2\right],
\label{bazar00}
\EEA
where $\ep^{\rm T}=(\ep_1,...,\ep_n)$ is the vector of eigenvalues of
$\hH$, $Z$ is the partition sum defined in (\ref{gibbs}), and where
$\ep_{\rm min}$ and $\ep_{\rm max}$ are the minimal and maximal ones
among $(\ep_1,...,\ep_n)$.

Since all the eigenvalues $\nu$ of the product of a double-stochastic
matrix to its transpose satisfy $0\leq \nu\leq 1$ 
\footnote{\label{fkuram} For any double-stochastic matrix $C_{ik}$, consider
the matrix $CC^{\rm T}$, where $C^{\rm T}$ is the transpose of $C$,
and let $a_i$ be an eigenvector of $CC^{\rm T}$ corresponding to a
(necessarily non-negative) eigenvalue $\nu$:
$\sum_{k,l=1}^nC_{ik}C_{lk}a_l=\nu a_i$.  One has
$|\sum_{k=1}^nC_{ik}a_k|=|\nu a_i|=\nu \,|a_i|\leq
\sum_{k,l=1}^nC_{ik}C_{lk}|a_l|$, and then
$\nu\sum_{i=1}^n|a_k|\leq\sum_{k=1}^n|a_k|$, that is, $\nu\leq 1$.}, one
has
\BEA
\ep^{\rm T}(1-CC^{\rm T})\ep\geq 0.
\EEA

Thus $\langle e^{-\beta w}\rangle$ is strictly smaller than unity.  As compared
to our discussion of the classical situation in section
\ref{messages}, the result $\langle e^{-\beta w}\rangle<1$ does not in
general permit to draw quantum analogs of the classical features ${\it
b}$ (active realizations) and ${\it c}$ (dispersion at high $T$) in
section \ref{messages}.

\section{Comparison with other approaches.}
\label{others}

In the present section we study two approaches known in literature. The
purpose is to understand whether they have the proper physical meaning
for describing fluctuations of work. Since they both allow to generalize
the classical BK equality (though in different ways), the adoption of
either of them will mean |as we discuss in detail below| that there is
no major qualitative difference in behavior of quantum and classical
fluctuations of work. It should perhaps be stressed that our concern is
the applicability of these approaches for describing fluctuations of
work under conditions formulated in the Introduction; their usefulness 
for other purposes is neither discussed, nor criticized. 

\subsection{Observable of work.}
\label{others1}

Recall
from definitions (\ref{quantwork}, \ref{deldef}) that for
any initial ensemble described by $\hrho$, the average of $\Del$ is
equal to the work done on the corresponding ensemble.

The approach goes on by stating \cite{bk,lindblad,yukawa,sasaki} that
the operator $\Del$ is the ``observable of work'' in the standard
sense of quantum observables \footnote{Once $\Del$ is given an
independent meaning as a quantum observable, there arises a 
question on its measurability, since the standard theories of 
quantum measurements, see e.g. \cite{willi,peres},
operate in Schr\"odinger representation. We shall not 
pursue this problem here, but rather take as working
hypothesis that this measurement can be carried out.}, 
e.g., the quantity ${\rm
tr}[\hrho\,\Del^2]-[{\rm tr}(\hrho\,\Del)]^2$ is to be interpreted as
the dispersion of work for any $\hrho$. However, while ${\rm
tr}\,[\Del\,\hrho]$ happens to be equal to the average energy lost by
the work source $\W$, simply due to conservation of the average energy
during the system-work-source interaction, this alone is, of course,
not sufficient to regard $\Del$ as an operator of work. In fact, such
an interpretation relies on the analogy between the definition
(\ref{deldef}) of $\Del$ and the classical expression (\ref{8-}) for
energy difference.  Such analogies are very widespread in general, and
once it is accepted that $\Del$ represents the proper energy
difference operator, the extension of its interpretation toward
operator of work seems rather natural.

Let us however recall from our discussion in the Introduction that we
expect for a proper approach to fluctuations of work to apply in
arbitrary non-equilibrium situation.  It is now possible to argue that
in general $\Del$ does not have the proper meaning of energy
difference operator, let alone its meaning as the operator of work.

Let the ensemble $\E(\hrho)$ have a density matrix
$\hrho(0)=|0\rangle\langle 0|$, such that $|0\rangle$
is an eigenstate of $\Del\equiv\hU^\dagger_\tau\,\hH\,\hU_\tau-\hH$
with eigenvalue zero: 
\BEA
\label{tut}
\Del|0\rangle=0. 
\EEA
Recall that $\hU^\dagger_\tau\,\hH(\tau)\,\hU_\tau$ is the Hamiltonian in
the Heisenberg representation a time $\tau$, while the Schr\"odinger
picture relation $\hH(\tau)=\hH$ is due to the assumed cyclic feature
of the process.

In general,
\BEA
[\hU^\dagger_\tau\,\hH\,\hU_\tau,\hH]\not=
0, 
\label{nono}
\EEA
so that $|0\rangle$ is neither an eigenstate of
$\hU^\dagger_\tau\,\hH\,\hU_\tau$, nor an eigenstate of
$\hH$.

According to quantum mechanics, Esq.~(\ref{tut}) should be interpreted
as follows: the operator $\Del$ has on the ensemble
$\E(|0\rangle\langle 0|)$ a definite value equal to zero, that
is, if it is interpreted as the operator of energy change, then for
{\it all single systems} from $\E(|0\rangle\langle 0|)$ the energy does not
change during this thermally isolated process.

There are however concrete examples 
|see Appendix \ref{sanik}|
showing that (\ref{tut}) can be consistent with
\BEA
\label{tsakat}
\left\langle 0\left| \,\left[
\hU^\dagger_\tau\,\hH\,\hU_\tau
\right]^m
\,  \right| 0\right\rangle\not =
\langle 0| \hH^m  | 0\rangle,\quad {\rm for}\quad m>2.
\EEA
This shows that the energy does change, since some of its moments do.
In other words, the interpretation of $\Del$ as the energy difference
operator is in general unsupportable.  Note that the non-commutativity
feature as expressed by (\ref{nono}) is essential for this conclusion.

\subsubsection{Restricted interpretation of $\Del$.}
\label{cricri}

A more restricted interpretation of $\Del$ can be given in the light
of the definition of fluctuations of work discussed in section
\ref{quant}. This will also show that if $\rho$ commutes with $\Del$
(a semiclassical assumption),
our approach is consistent with that of the observable of work.

Let the eigenresolution of $\Del$ be
\BEA
\label{oomm}
\Del=\sum_{k=1}^n\omega_k|\omega_k\rangle\langle\omega_k|.
\EEA

Note that for $\Del$ to have the meaning of the operator of work it is
necessary that {\it i)} its eigenvalues $\{\omega_k\}_{i=1}^n$ have
the meaning of work by themselves, i.e., $\omega_k$ should have both
the meaning of average energy lost by the work source $\W$ and the
average energy gained by a quantum ensemble, as we discussed in
section \ref{qqff}; {\it ii)} probabilities of these realizations of
work done on the initial ensemble $\E(\hrho)$ should be given as
$\{\,\langle\omega_k|\hrho|\omega_k\rangle\,\}_{k=1}^n$.  

Now, if $\hrho$ and $\Del$ commute,
\BEA
[\hrho,\Del]=0,
\label{dzen}
\EEA
then their eigenvectors can be chosen the same, and,
by measuring $\Del$,
$\hrho=\sum_{k=1}^np_k |\omega_k\rangle\langle\omega_k|$
can be separated into subensembles
$\{\E(|\omega_k\rangle\langle\omega_k|)\}_{k=1}^n$
with probabilities $p_k=\langle\omega_k|\hrho|\omega_k\rangle$. 
The average work done on each subensemble
$\E(|\omega_k\rangle\langle\omega_k|)$ is then equal to $\omega_k
=\langle\om_k|\Del|\om_k\rangle$,
and one can admit the restricted interpretation of $\Del$ as an
operator of work.

Conversely, if $\hrho$ can be separated into subensembles, 
\BEA
\label{den}
\hrho=\sum_{k=1}^n\lambda_k
|\psi_k\rangle\langle\psi_k|, 
\EEA
and if each of them is let to
interact with the work source $\W$ such that 
\BEA
\label{mao}
\omega_k=\langle\psi_k|\Del|\psi_k\rangle,\quad
\lambda_k=\langle\omega_k|\hrho|\omega_k\rangle,
\EEA
then three conditions
(\ref{den}, \ref{mao}, \ref{oomm}) imply commutation (\ref{dzen}).

To show this we proceed in a slightly indirect way, which is useful by
itself.  It can be noted that the dispersion
\BEA
\label{espanol}
{\rm tr}[\hrho\,\Del^2]-[{\rm tr}(\hrho\,\Del)]^2= 
\sum_{k=1}^n
\langle\omega_k|\hrho|\omega_k\rangle \, (\omega_k-W)^2
\EEA
of the operator $\Del$ provides an upper bound for the maximal dispersion
$\del_{\rm max}$ of work given by Eq.~(\ref{dard1}):
\BEA
\label{camaradas}
&&{\rm tr}(\hrho\Del^2)-W^2
-\del_{\rm max}\nonumber\\
&&=\half\sum_{i,k=1}^n
\frac{(p_i-p_k)^2}{p_i+p_k}\,|\langle p_k|\Del | p_i\rangle|^2\geq 0.
\EEA
The equality in the RHS of (\ref{camaradas}) is realized only if
$\hrho$ and $\Del$ commute, that is, either $\langle p_k|\Del
| p_i\rangle$ is zero for $i\not =k$, or for some pair $i\not =k$ one
has $\langle p_k|\Del | p_i\rangle\not =0$, but the corresponding
eigenvalues of $\hrho$ are degenerate: $p_i=p_k$. Thus $\del_{\rm
max}$ can be equal to ${\rm tr}(\hrho\Del^2)-W^2$ only if 
$[\hrho,\Del]=0$.

Now note that if Eq.~(\ref{den}, \ref{mao}, \ref{oomm}) are 
assumed to be valid, they
imply ${\rm tr}(\hrho\Del^2)-W^2 -\del_{\rm max}\leq 0$ simply due
to the definition of the maximal dispersion.  
This is consistent with Eq.~(\ref{camaradas}) only for
${\rm tr}(\hrho\Del^2)-W^2 -\del_{\rm max}= 0$, which implies
$[\hrho,\Del]=0$ as we saw above. We conclude that (\ref{den},
\ref{mao}, \ref{oomm}) imply (\ref{dzen}), as was promised.

Thus, when $[\hrho,\Del]\not=0$, $\Del$ does not qualify as the
operator of work even in the restricted sense.  We also conclude that
though the approach does predict an upper bound for $\del$, this
bound is not reachable
\footnote{\label{sup} Note that the difference ${\rm tr}(\hrho\Del^2)-W^2
  -\delta W^2=\sum_{\alpha=1}^N\lambda_\alpha \left(\,
  \langle\psi_\alpha|\Del^2|\psi_\alpha\rangle-
  \langle\psi_\alpha|\Del|\psi_\alpha\rangle^2\,\right)\geq 0 $ is by
  itself always non-negative for any separation of $\hrho$ into
  subensembles.}.

\subsubsection{On a generalization of the classical BK equality.}

Though $\Del$ does not have the meaning of the operator of work 
| except in the restricted sense and under condition (\ref{dzen}) |
there is an operator generalization of 
Eqs.~(\ref{gort}, \ref{dodosh}) which
was proposed by Bochkov and Kuzovlev in \cite{bk,bkphysica}:
\BEA
\label{q1}
\frac{{\rm tr}\, e^{-\beta\Del-\beta\hH}}{Z}=
\left\langle\, \overrightarrow{\exp}\left[
-\int_0^\beta\d s\,e^{-s\hH}\Del \,e^{s\hH}
\right]\,\right\rangle\\
\equiv\tr\,\left(\,\overrightarrow{\exp}\left[
-\int_0^\beta\d s\,e^{-s\hH}\Del \,e^{s\hH}
\right]\,\hrho\,\right)=1.
\label{q11}
\EEA
We recall its derivation in Appendix \ref{bochka}
\footnote{
For the equilibrium ensemble (\ref{gibbs}), the
Thomson formulation of the second law 
can be derived from (\ref{q1}, \ref{q11}) upon the application of
the Peierls-Bogoliubov inequality (recalled in Appendix \ref{bochka}): 
$e^{-\beta\,\tr[\hrho\,\Del]}\leq \frac{1}{Z}\,{\rm tr}\,
e^{-\beta\Del-\beta\hH}=1$.  
From this it follows once again that 
$W={\rm tr}\,[\Del\,\hrho] \geq 0$.}. 
A similar relation was derived in \cite{yukawa}.

Let us work out some consequences of (\ref{q11}).  As compared to the
classic case, the matters are complicated by the presence of
anti-time-ordering and the integral $\int_0^\beta$ in (\ref{q1},
\ref{q11}). If one would insist on not having them, then the equality
(\ref{q1}, \ref{q11}) can still be converted into an inequality. By
applying Thompson-Golden inequality \cite{gt}
\footnote{\label{gt1} Thompson-Golden inequality is a particular consequence
of the following submajorization relation
$\lambda\left(e^{\hat{A}+\hat{B}}\right) \prec_w
\lambda\left(e^{\hat{A}/2}e^{\hat{B}}e^{\hat{A}/2}\right)$, where
$\lambda\left(\hat{A}\right)$ is the eigenvalue vector of a hermitian
operator $\hat{A}$; see ~\cite{major3} for more details.},
$\tr\,[\,e^{\hat{A}}\,e^{\hat{B}}\,]\ge\tr\,e^{\hat{A}+\hat{B}}$,
valid for any hermitian operators $\hat{A}$ and $\hat{B}$ (the
equality sign is realized here if and only if $[\hat{A},\hat{B}]=0$),
one gets
\BEA 
\label{dzuk}
\langle e^{-\beta\Del}\rangle
\equiv
{\rm tr}\,[\, \hrho\,e^{-\beta\Del}\,]\nonumber\\
=\sum_{k=1}^n\langle\omega_k|\hrho|\omega_k\rangle\,
e^{-\beta\omega_k}
\geq 
\frac{1}{Z}\,{\rm tr}\, e^{-\beta\Del-\beta\hH}=1,
\EEA
where $|\omega_k\rangle$ and $\omega_k$ are eigenvectors and eigenvalues of
$\Del$ as defined by (\ref{oomm}).

If now we could interpret $\Del$ as the operator of work, that is, if
the eigenvalues $\omega_k$ of $\Del$ would have the meaning of work by
themselves, we would note that $\langle\omega_k|\hrho|\omega_k\rangle$
is the probability of observing the eigenvalue $\omega_k$ upon the
measurement of $\Del$ on the state $\hrho$, and then Eq.~(\ref{dzuk})
would allow us to study fluctuations of work exactly in the way we did
in section \ref{messages} for the classical situation.  We would then
draw the same general conclusions, and the fact that (\ref{dzuk}) is
an inequality will only {\it strengthen} these conclusions as compared to
the classical situation. However, as we saw above, it is impossible to
identify $\Del$ with the operator work, and thus fluctuations of work
cannot be studied on the base of (\ref{dzuk}), except for
$[\hrho,\Del]=0$, where Eqs.~(\ref{q1}, \ref{q11}, \ref{dzuk}) reduce
to the usual (essentially classical) BK equality.

\subsection{On the approach based
on two-time measurements of energy.}
\label{twotimeapproach}

Yet another, different approach to fluctuations of work and extension
of the classical BK equality was proposed in
Refs.~\cite{kurchan,tasaki,mukamel}.  We shall present it in a more
extended form, since it is necessary for the understanding of its
proper physical meaning. On the other hand, in order do not dwell into
unnecessary technical details, we shall assume that the spectrum of the
Hamiltonian $\hH$ is non-degenerate (compare with (\ref{dunaj}))
\BEA
\label{desna}
\ep_1< \ep_2<...< \ep_n.
\EEA

At the time $t=0$ one measures energy (corresponding to the operator
$\hH$) for the ensemble described by the gibbsian density matrix
(\ref{gibbs}). The probability to get an eigenvalue $\eps_l$ of $\hH$
is seen from (\ref{resolH}) to be
\BEA
\label{1*}
\Pl&=&\langle\ep_l| \hrho| \ep_l\rangle,\\
   &=&p_l.
\label{2*}
\EEA
Eq.~(\ref{1*}) is the general quantum formula (Born's rule),
while Eq.~(\ref{2*}) follows from the Gibbsian form 
(\ref{gibbs}, \ref{resolrho}, \ref{pk}) 
of $\hrho$. The symbol ${\cal M}_0$ in (\ref{1*})
reminds that the probability is conditional and refers to the
measurement of $\hH$ done at $t=0$. The necessity of such explicit
notations will be seen below. Formally it is always allowed, since
{\it any} probability is conditional.

According to Wigner's formula for multi-time probabilities 
in quantum mechanics \cite{wigner}, the
subsequent measurement of energy at the time $\tau$ | represented by
the same Hamiltonian $\hH$ due to the cyclic feature of the considered
process | will then produce a result $\eps_k$ with the conditional
probability
\BEA
\Pkl=
|\langle\eps_k|\hU_\tau|\eps_l\rangle|^2.
\label{dudu}
\EEA 
There three conditionals for the probability in the LHS of
(\ref{dudu}): ${\cal M}_0$ and ${\cal M}_\tau$ stand for the
measurements done at $t=0$ and $t=\tau>0$, while the index $l$
indicate on the result $\ep_l$ got during the first measurement.  The
meaning of (\ref{dudu}) is that the ensemble of systems which during
the first measurement at $t=0$ produced the result $\ep_l$, is
described for $t>0$ by $|\ep_l\rangle\langle\ep_l|$.  The members of
this ensemble couple to the work source $\W$, the state evolves to
$\hU_\tau\,|\ep_l\rangle\langle\ep_l|\,\hU_\tau^\dagger$ at the time
$t=\tau$, and then is subjected to the second measurement.

Thus the total probability for having the result $\eps_l$ at the
moment $t=0$ {\it and} the result $\eps_k$ at $t=\tau$ is given by
\BEA
\label{3*}
\Pkandl &=& p(l/{\cal M}_0,{\cal M}_\tau)\,\Pkl~~~~\\
&=& \Pl\,\Pkl.
\label{fff}
\EEA
When passing from (\ref{3*}) to (\ref{fff}), we used the obvious relation
$p(l|{\cal M}_0,{\cal M}_\tau)= \Pl$ (no dependence on the future).

It is to be noted that 
\BEA
p(k|{\cal M}_0,{\cal M}_\tau)=
\sum_{l=1}^n\Pkandl\nonumber\\
=\sum_l p_l  
\langle\eps_k|\hU_\tau\,|\ep_l\rangle\langle\ep_l|
\,U^\dagger_\tau|\eps_k\rangle,
\EEA
that is, the probability to have the result $\ep_k$ at the second
measurement is for a general initial density matrix $\hrho$ not
equal to 
\BEA
p(k|{\cal M}_\tau)=
\langle\eps_k|\hU_\tau\,\hrho\,U^\dagger_\tau|\eps_k\rangle ,
\EEA
which is the probability to get the result $k$ in a different context,
where no first measurement was done. 
Such an equality is valid, though, if $\hrho$ commutes with $\hH$, 
which is the case with the Gibbsian density matrix (\ref{gibbs}).
Let us first restrict our attention to this case.
One notes from (\ref{dudu}) the double-stochastic feature of
$\Pkl$:
\BEA
\label{blu}
\sum_{k=1}^n\Pkl=\sum_{l=1}^n\Pkl=1,
\EEA
and calculates using (\ref{resolrho}, \ref{pk}, \ref{2*}, \ref{blu}):
\BEA
&&\langle e^{-\beta(\eps_k-\eps_l)}\rangle_{0,\tau}\nonumber\\
&&\equiv
\sum_{k,l=1}^n\Pl\,\Pkl\,e^{-\beta(\eps_k-\eps_l)}\nonumber\\
&&=\frac{1}{Z}\sum_{k,l=1}^n\,\Pkl\,e^{-\beta\eps_k}=1.
\label{q2}
\EEA

This is the equality got in Refs.~\cite{kurchan,tasaki,mukamel}
as a generalization of the classic BK equality.

Note that for the density matrix (\ref{gibbs})
the average 
\BEA
\label{gi}
\sum_{k,l=1}^n\Pl\,\Pkl\,(\eps_k-\eps_l) =W,
\EEA
is equal to the work as defined by (\ref{quantwork}).  The statement
of the second law, $W\geq 0$, can once again be deduced from (\ref{q2})
by employing convexity of the exponent.

\subsubsection{Critique of the approach.}
\label{cricricri}

Would now we be able to associate the work with a random variable
having realizations $\{\eps_k-\eps_l\}_{k,l=1}^n$ and the
corresponding probabilities $\{\Pkl\}_{k,l=1}^n$, it would be possible
to study fluctuations of work on the base of Eq.~(\ref{q2}), and to
draw essentially the same conclusions as we did in section
\ref{messages} for the classical case.  It is, however, not difficult
to see that the same criticisms we brought in section \ref{others1}
with respect to the ``observable of work'' applies here too.

Keeping in mind our discussion after Eq.~(\ref{dudu}), note that
if the ensemble initially described by $|\ep_l\rangle\langle\ep_l|$
couples to the work source $\W$, its mechanical degree of freedom
looses at the time $t=\tau$ the energy
\BEA
\label{trombon}
\tr \left(\Del\,|\ep_l\rangle\langle\ep_l|\right)=
\tr\left( \hH\,\hU_\tau\,|\ep_l\rangle\langle\ep_l|\,\hU^\dagger_\tau
\right)-\ep_l. 
\EEA
Since the final density matrix
$\hU_\tau\,|\ep_l\rangle\langle\ep_l|\,\hU^\dagger_\tau$ need not
commute with $\hH$, the energy need not any definite value at that
time, and Eq.~(\ref{trombon}) does in general not reduce to
$\ep_k-\ep_l$ with any $k$.  Such a reduction takes place when
\BEA
\label{se}
\tr\left( \hH\,\hU_\tau\,|\ep_l\rangle\langle\ep_l|\,\hU^\dagger_\tau
\right)=
\sum_{k=1}^nC_{kl}\ep_k=\ep_{{\bf \pi}(l)},\\ 
{\rm for} \quad l=1,...,n,
\EEA
where $C_{kl}$ is defined via (\ref{dudu1}), and
where $({\bf \pi}(1),...,{\bf \pi}(n)\,)$ is some permutation of 
the sequence $(1,...,n)$. Eq.~(\ref{se}) can now be re-written as
\BEA
\label{aa}
\sum_{k=1}^n\widetilde{C}_{kl}(\ep_k-\ep_{l})=0, 
\EEA
where the matrix $\widetilde{C}=C\,\Pi$ the product of $C$ and the
corresponding permutation matrix $\Pi$, and where we noted that once
the matrices $C$ and $\Pi$ are double-stochastic (see (\ref{ds}) for
definition), so is $\widetilde{C}$.  Note with help of Eq.~(\ref{desna})
that for $l=n$ all terms with $k\not =n$ in (\ref{aa}) are negative
unless $\widetilde{C}_{k\not =n}=0$, which via the double-stochastic
feature of $\widetilde{C}$ implies: $\widetilde{C}_{n\not =k}=0$ and
$\widetilde{C}_{n n}=1$.  Continuing along the same lines for $l<n$,
one gets that (\ref{se}) can take place only when $\widetilde{C}$
reduces to unity matrix, or, equivalently, $C$ reduces to a
permutation matrix:
\BEA
\hU_\tau\,|\ep_l\rangle\langle\ep_l|\,\hU^\dagger_\tau=
|\ep_{\pi(l)}\rangle\langle\ep_{\pi(l)}|.
\label{damask}
\EEA

Thus, in general it is (\ref{trombon}) and not $\ep_k-\ep_l$ that
can be interpreted as the work for this single realization occuring
with probability $p_l$, and this is precisely the point from which we
departed in section \ref{quant}.

It is also straightforward to see that the approach does not apply
out of equilibrium. The reasons for this are more straightforward
than for the previous approach.

Recall from the Introduction that the proper definition of
fluctuations of work is expected to apply to any non-equilibrium
initial ensemble $\E(\hat{\sigma})$ with density matrix $\hat{\sigma}$
not commuting with $\hH$:
\BEA
\label{zv}
[\hat{\sigma},\hH]\not =0.
\EEA
In particular, the work averaged over those fluctuations should be
equal to the one done on the ensemble.

The present approach is generalized uniquely for arbitrary initial
state: the definitions of $\Pl$ and $\Pkl$ in Eqs.~(\ref{dudu},
\ref{1*}) remains unaltered: one substitutes there $\hat{\sigma}$
instead of $\hrho$.

It is now straightforward to see from (\ref{zv}) that due to 
non-diagonal terms in $\hat{\sigma}$, the average
$\sum_{k.l=1}^n\Pl\,\Pkl\,(\eps_k-\eps_l)$ is not equal to the work
$\tr(\hat{\sigma}\Del)$ done on the overall ensemble:
\BEA
\tr (\hat{\sigma}\Del)
-\sum_{k.l=1}^n\Pl\,\Pkl\,(\eps_k-\eps_l)\nonumber\\
=\tr\left[
\hU^\dagger_\tau\hH\hU\left(\,
\hsigma-|\ep_l\rangle\langle\ep_l|\hsigma|\ep_l\rangle\langle\ep_l|
\,\right)
\right]\not =0.~~
\EEA

\subsubsection{
The approaches based on the ``observable of work'' and
on two-time measurements of energy are different. }

This is seen already by comparing Eq.~(\ref{dzuk}) with 
(\ref{q2}). Still we want to understand this difference in more detail.
More precisely, though for the initial density matrix commuting
with $\hH$, the first and the 
second moments generated by the two approaches are
equal:
\BEA
{\rm tr}[\hrho\Del^p]=
\sum_{k,l=1}^n\Pl\,\Pkl\,(\eps_k-\eps_l)^p,\\
p=0,1,2,~~~~~
\EEA
already the third moments are in general different, even for
$[\hrho,\hH]=0$. Indeed,
assuming validity of the latter condition, one gets
\BEA
{\rm tr}[\hrho\Del^3]-
\sum_{k,l=1}^n\Pl\,\Pkl\,(\eps_k-\eps_l)^3\nonumber\\
=\tr\left(\Del\,[\Del,\hrho]\,\hH\right)
=\tr\left(\hrho\,[\hH,\Del]\,\Del\right).
\label{gaga}
\EEA

The RHS of (\ref{gaga}) vanishes if 
$[\hrho,\Del]=0$, or $[\hH,\Del]=0$,
in addition to $[\hrho,\hH]=0$.  

Note as an illustration that for
the two-level example of section \ref{2l} the RHS of (\ref{gaga})
reads:
\BEA
\label{gavgav}
\tr\left(\hrho\,[\hH,\Del]\,\Del
\right)=(p_1-p_2)(\ep_1-\ep_2)|\chi|^2,
\EEA
where $\Del$ is given by (\ref{dada}), and 
where $p_k$ and $\ep_k$ are eigenvalues of $\hrho$ and $\hH$, respectively.
For the Gibbsian density matrix $\hrho$, the RHS of (\ref{gavgav})
has negative sign.

Finally note that differences between the two approaches were recently
studied in ~\cite{sasaki} in a different context.

\subsection{Summary.}

We have discussed two approaches known in literature, and argued that in
the proper quantum domain, though being related to energy, they do not
describe fluctuations of work.  Work is a rather particular form of
energy having several specific features we discussed in section
\ref{workclass}.  These two approaches allow different generalizations of the
classical BK equality and this makes them operationally close to the
classical situation. Still this resemblance is superficial, since, as
we argued, it is not ensured by the two approaches that realizations
of the claimed random quantity/operator of work have themselves the
physical meaning of work.

\section{Conclusion.}
\label{conclusion}

The second law has a statistical character as it is both formulated
and valid for ensembles of identically prepared systems. It is
therefore of interest to investigate this statistical aspect in more
detail. For the entropic formulation of the second law, this analysis
is by now a standard chapter of statistical thermodynamics 
~\cite{epstein,tolman,landau}.

In the present paper we studied how Thomson's formulation of the
second law: no work from an equilibrium ensemble by a cyclic process,
emerges through averaging over fluctuations of work in the quantum
situation.  It will be useful at this moment to recall the special
role of Thomson's formulation, and then to proceed with concluding
remarks on our results.

\subsubsection{The main features of Thomson's formulation of the second law.}

\paragraph{}
The formulation uses the concept of work which is unambiguously defined
both conceptually and operationally, both in and out of equilibrium.
In this respect work is contrasting to entropy, which is well-defined
only in (nearly) equilibrium states of macroscopic systems.

\paragraph{}
Thomson's formulation is valid for any {\it finite}
\footnote{
In this context one sometimes hears that the second law must refer to
macroscopic systems, and there is no sense in applying it for finite
systems. This opinion is not correct, as instanced by Thomson's
formulation. Would it not be valid for a finite system coupled to work
sources, the very its application to macroscopic systems will be
endangered, because the initial Gibbsian ensemble (\ref{gibbs}) is
prepared under weak interaction with an equilibrium thermal bath, see
e.g. \cite{balian}: any cycle violating the formulation for a finite
system can be repeated to achieve a violation for the bath.}
or {\it infinite} \cite{woron}, quantum or classical system interacting
with macroscopic sources of work.  In this context one notes that not
all formulations of the second law have such a universal regime of
validity. While all formulations are equivalent in the standard
thermodynamical domain, that is, for (nearly) equilibrium states of
macroscopic systems, some of them have definite limits when considered
for finite systems~ \cite{an} or for low temperatures (quantum domain)
~\cite{AN,NA}.

\paragraph{}
In its literal form Thomson's formulation does not imply any
irreversibility, since the dynamics of the system coupled to work
source is unitary and thus formally reversible: if some work was put
into the initially equilibrium system it can in principle be extracted
back. The irreversibility with respect to work transfer comes into
existence when one takes into account that in practice no work source
can interact with all possible degrees of freedom.  In particular, if
the system was subjected to a thermal bath after it had interacted
with the work source, it relaxes back to its Gibbsian state and the
work which had been put into it cannot be recovered by {\it any}
work-source acting on the system only (a similar argument is presented
in \cite{lindblad}, chapter 5). Thus the relation between Thomson's
formulation and irreversibility is seen clearly at least in the
simplest situation.

\paragraph{}
It should perhaps be stressed that Thomson's formulation does not
refer to all aspects usually associated with the second law, e.g., by
itself it does not explain how a subsystem of a proper macroscopic
system (thermal bath) relaxes toward a Gibbsian equilibrium state
(though on the basis of Thomson's formulation it is still possible to
argue that -- under certain assumptions -- the Gibbsian state is the
only one which forbids work extraction via {\it any} cyclic thermally
isolated process \cite{woron,lenard}). The property of relaxation toward
a Gibbsian state is to be viewed as an independent item of statistical
physics; its standard classical understanding was reshaped in
literature various times; see, e.g., ~\cite{loinger,tasakiprl,AN,NA}.

\subsubsection{What appeared to be 
problematic in defining fluctuations of work in the
quantum situation.}

As we saw in section \ref{others}, due to non-commutativity of various
quantum observables, there are different quantum quantities which, in
the classical limit, coincide with the random quantity work. Thus, as often,
classical reasoning alone is of no help for defining fluctuations of
work. 

One therefore has first to state what basic features the
fluctuations of work are going to have, as we did in section 
\ref{intro}, and then to recognize that the work is an essentially
mechanical, classical quantity | in spite of the fact that 
it can be added or
extracted from a quantum system | since it is an energy transferred to
macroscopic degrees of freedom of the work-source that, at least in
principle, should be retransferrable to other classical sources.

Once this feature was recognized, the definition of fluctuations of
work we presented in section \ref{quant} is straightforward.

\subsubsection{What is similar and what is different
in classical and quantum definitions
of fluctuations of work?}

In both situations the definition of work as a random quantity employs
the same idea: the initial homogeneous ensemble of identically
prepared systems is separated into irreducible (homogeneous)
subensembles. Both in quantum and classics these irreducible
subensembles are described maximally completely. In classics they
correspond to a trivial subensemble of identical copies of the same
system (so that within a subensemble no fluctuations are present
whatsoever), and they are described via phase-space points and
trajectories. In quantum mechanics these subensembles, described by
pure density matrices (wave functions), provide definite
(non-fluctuating) values for the largest possible, {\it but non-exhaustive},
set of observables.

In classics the irreducible subensembles of the initial ensemble
obviously exist {\it a priori}, that is, without need of any
measurement.  In the quantum situation the very structure of
subensembles does depend on the measurement applied for the actual
separation, or, in other words for the preparation of an ihomogeneous
ensemble. As the main consequence, the separation of a mixed ensemble
is not unique, and thus the random quantity work is {\it contextual}
in the quantum situation. It is therefore to be recalled that in the
quantum situation the definition of fluctuations of work always needs
this initial {\it preparational} measurement, a step which is absent
in classics.

In the second step, systems from each irreducible subensemble interact
with the same macroscopic source which does on them the same thermally
isolated process. Realizations of the random quantity work are then
defined as the average energy increase of the work-source when
interacting with each subensemble, while the probability of each
realization is given by the weight of the corresponding subensemble in
the initial mixed ensemble. 

In this way the full physical meaning of work is kept, and 
the approach can be applied to any non-equilibrium initial state.

\subsubsection{Dispersion of work.}

The most direct quantity which characterizes fluctuations of work is
the dispersion of work we studied in section
\ref{dispersion}. Although the work is a contextual random quantity
and depends on the measurement that was done to separate the initial
mixed ensemble into pure subensembles, one can define two reasonable
quantities |maximal dispersion and the dispersion for a randomly
chosen separation on the initial ensemble| that depend solely on the
internal features of the system, that is, on its initial state and its
time-dependent Hamiltonian.  These quantities were calculated
explicitly for any finite quantum system and studied in section
\ref{dispersion}.

\subsubsection{Non-existence of the direct generalization of classic BK
equality.}

In the classical situation fluctuations of work in an initially
equilibrium state are controlled by the BK equality \cite{bk,jar}.
This equality allows to draw a number of model-independent statements
on fluctuations of work which we summarized in section \ref{messages}.
In contrast, the {\it direct} generalization of BK equality to the
quantum domain |the one which would allow to draw similar qualitative
conclusion on fluctuations of work| does not exist; see section
\ref{no}. As we discussed in detail in section \ref{others}, there are
quantum generalizations of BK equality, but they refer to quantities
which describe fluctuations of work only if some classical features
are present, e.g. those implied by Eq.~(\ref{dzen}). As the main
consequence, fluctuations of work in the quantum situation can have
features which are impossible in classics, e.g., (inter-subensemble)
fluctuations can be absent completely.

\acknowledgments It is a pleasure to thank Roger Balian for inspiring
discussions. We thank Claudia Pombo for critical remarks and for
stressing the importance of understanding of the concept of work.

The work of A.E. A is part of the research programme of the Stichting voor 
Fundamenteel Onderzoek der Materie (FOM, financially supported by 
the Nederlandse Organisatie voor Wetenschappelijk Onderzoek (NWO)).

\appendix

\section{Derivation of Eq.~(\ref{tutmos2}).}
\label{mitri}

Here we recall from \cite{mitri} a generalization of Cauchy inequality
used in Eq.~(\ref{tutmos2}).

Denote by $\Gamma=(x,p)$ the phase space point. Assume that al the integrals
over the phase-space used below are finite. The desired inequality
reads: if $a(\Gamma)$, $b(\Gamma)$, $x(\Gamma)$ are some functions
satisfying
\BEA
\int \d \Gamma\, a(\Gamma)\,x(\Gamma)=0,\quad
\int \d \Gamma\, b(\Gamma)\,x(\Gamma)=1,
\EEA
then
\BEA
&&\int \d \Gamma\, x^2(\Gamma)\nonumber\\ 
&&\geq
\frac{\int \d \Gamma\, a^2(\Gamma)}
{\int \d \Gamma\, a^2(\Gamma)\int \d \Gamma\, b^2(\Gamma)-
\left[
\int \d \Gamma\, a(\Gamma)\,b(\Gamma)
\right]^2}.
\label{dadada}
\EEA
To prove this, define
\BEA
\label{tkhra1}
A=\int \d \Gamma\, a^2(\Gamma),\\
\label{tkhra2}
B=\int \d \Gamma\, b^2(\Gamma),\\
\label{tkhra3}
C=\int \d \Gamma\, a(\Gamma)\,b(\Gamma),\\
\label{tkhra4}
y(\Gamma)=\frac{A\,b(\Gamma)-C\,a(\Gamma)}{AB-C^2},
\EEA
and note that
\BEA
\label{bizon}
\int \d \Gamma\, x^2(\Gamma)\geq
\int \d \Gamma\, y^2(\Gamma),
\EEA
due to
\BEA
\int \d \Gamma\, x(\Gamma)\,y(\Gamma)=
\int \d \Gamma\, y^2(\Gamma),
\EEA
which is valid
by constructions (\ref{tkhra1}--\ref{tkhra4}). Eq.~(\ref{dadada})
follows from (\ref{bizon}). To get from here Eq.~(\ref{tutmos2}) one
identifies: $x(\Gamma)=\sqrt{{\cal P}(\Gamma)}$, 
$b=\sqrt{{\cal P}(\Gamma)}\,e^{-\beta w(\Gamma)}$,
$a=\sqrt{{\cal P}(\Gamma)}\,(f(\Gamma)-\langle f\rangle)$.

\section{Derivation of Eq.~(\ref{bazar}).}
\label{secul}

Let $f(x)$ is a smooth function, $\{x_i\}_{i=1}^n$ are $n$ points, and
\BEA
\bar{x}=\sum_{k=1}^n\lambda_kx_k,\quad
\lambda_k\geq 0,\quad
\sum_{k=1}^n\lambda_k=1.
\EEA

Apply the incomplete Taylor expansion to $f(x_i)$:
\BEA
\label{inco}
f(x_i)=f\left(\bar{x}\right)
+f'(\bar{x})\,\left(
x_i-\bar{x}
\right)
+\frac{f''(\xi_i)}{2}
\,\,\left(
x_i-\bar{x}
\right)^2,
\EEA
where $\xi_i$ lies between $x_i$ and $\bar{x}$. Denote by
$x_{\rm max}$ and $x_{\rm min}$ the maximal and the minimal 
numbers among $x_i$.
This implies $x_{\rm max}\geq \xi_k\geq x_{\rm min}$.
Now assume that $f''(x)$ is monotonically decaying:
\BEA
\label{borshomi2}
f''(x_{\rm max})\geq f''(\xi_i)\geq f''(x_{\rm min}).
\EEA
Using (\ref{inco},\ref{borshomi2}) one has
\BEA
\sum_{k=1}^n\lambda_kf(x_k)-f(\bar{x})=
\frac{1}{2}\sum_{k=1}^nf''(\xi_{k})
\lambda_k(x_k-\bar{x})^2,\\
\frac{f''(x_{\rm min})}{2}\sum_{k=1}^n\lambda_k(x_k-\bar{x})^2
\geq
\sum_{k=1}^n\lambda_kf(x_k)-f(\bar{x})
\nonumber\\
\geq
\frac{f''(x_{\rm max})}{2}\sum_{k=1}^n\lambda_k(x_k-\bar{x})^2.
\label{rinok}
\EEA
To derive Eq.~(\ref{bazar}), start from (\ref{shuka}), apply to it
Eq.~(\ref{rinok}) with a convex function $f(x)=e^{-\beta x}$, 
$\beta=1/T>0$, and identify $x_i=\ep_i$, $\lambda_k=C_{kl}$. The desired
Eq.~(\ref{bazar}) is recovered upon the summation over $l$.

\section{}
\label{f7}

Let $\tr (\hat{A}^2\hrho)=[\tr (\hat{A}\hrho)]^2$ for some hermitian
operator $\hat{A}$ and density matrix $\hrho$.  In the main text we
called such operators dispersionless with respect to the ensemble
described by the density matrix $\hrho$.

In Cauchy inequality $|\tr (\hat{A}\hat{B})|^2\leq \tr
(\hat{A}\hat{A}^\dagger) \,\tr (\hat{B}\hat{B}^\dagger)$, which is
valid for any operators $\hat{A}$ and $\hat{B}$, the equality is
realized for $\hat{A}=\alpha\hat{B}^\dagger$, where $\alpha$ is a
number. Thus 
\BEA
\left[\tr (\hat{A}\,\sqrt{\hrho}\,\sqrt{\hrho})\right]^2= \tr
(\hat{A}^2\hrho)\,\tr(\hrho) 
\EEA
implies 
\BEA
\label{c4}
\hat{A}\sqrt{\hrho} =\alpha\,\sqrt{\hrho}. 
\EEA
Now apply the eigenresolution $\sqrt{\hrho}=\sum_{k=1}^n
\sqrt{p_k}\,|\eps_k\rangle\langle\eps_k|$ into (\ref{c4}) and multiply
it from the right by $|p_m\rangle$ to obtain:
\BEA
\sqrt{p_m}\,\hat{A}\,|p_m\rangle
=\alpha\,\sqrt{p_m}\,|p_m\rangle.
\EEA
It is seen that either only one among the eigenvalues $p_k$'s is
non-zero and then the corresponding eigenvector is also an eigenvector
for $\hat{A}$, or, more generally, that $\hat{A}$ acts as $\propto
\hat{1}$ in the Hilbert space formed by eigenvectors of $\hrho$
corresponding to its non-zero eigenvalues.  In both cases the
measurement of $\hat{A}$ on the state $\hrho$ always produces definite
results.

Thus any operator $\hat{A}$ that is dispersionless on the density matrix 
$\hrho$ has to have the following block-diagonal matrix representation
\BEA
\hat{A}=\left(\begin{array}{rr}
\alpha\,\hat{1}_{k\times k} & ~~0 \\
0~~ & ~~\hat{B} \\
\end{array}\right),
\EEA
where $\alpha$ is a real number, $\hat{1}_{k\times k}$ is $k\times k$
unity matrix in the $k$-dimensional Hilbert space formed by 
eigenvectors corresponding to
non-zero eigenvalues of $\hrho$, and finally $\hat{B}$ is an arbitrary
$(n-k)\times (n-k)$ hermitian matrix. It has $(n-k)^2$ free parameters,
and another free parameter of $\hat{A}$ is coming with 
the real number $\alpha$. Thus,
$\hat{A}$ has 
\BEA
(n-k)^2+1,\nonumber
\EEA
free parameters.

Note finally that various operators that are dispersionless on a pure
density matrix need not be mutually commuting. As one of the simplest
examples take
\BEA
\hat{C}=\left(\begin{array}{rrr}
1 & 0 & 0 \\
0 & 0 & 0 \\
0 & 0 & 1 \\
\end{array}\right), ~
\hat{D}=\left(\begin{array}{rrr}
0 & 1 & 0 \\
1 & 0 & 0 \\
0 & 0 & \epsilon \\
\end{array}\right),~
|\psi\rangle=\left(\begin{array}{r}
0 \\
0 \\
1 \\
\end{array}\right),\nonumber
\EEA
where $\epsilon$ is real.
As seen, $\hat{C}|\psi\rangle=|\psi\rangle$ and 
$\hat{D}|\psi\rangle=\epsilon\,|\psi\rangle$, but $[\hat{C},\hat{D}]\not=0$.

\section{Relation between POVMs and projective measurements.}
\label{povm}

We outline how a POVM given by (\ref{comp}, \ref{durman}, \ref{yale})
can be connected with the usual projective measurements. 

Let the system $\S$ interact with another auxiliary system ${\cal
G}$. The initial states of $\S$ and ${\cal G}$ are, respectively,
$\hrho$ (living in a $n$-dimensional Hilbert space ${\cal H}$) and
$|g_1\rangle \langle g_1|$ living in a $N$-dimensional Hilbert space
${\cal H}'$. The initial state of the overall system $\S+{\cal G}$ is
$\hrho\otimes|g_1\rangle \langle g_1|$.

Select a fixed orthonormal base $\{|u_k\rangle\}_{k=1}^n$ in ${\cal
H}$.  Let as well $|g_1\rangle$ be a member of an orthonormal base
$\{|g_\alpha\rangle\}_{\alpha=1}^N$ in ${\cal H}'$.  Assume that the
above interaction is chosen such that the corresponding unitary
evolution operator $\hat{{\cal U}}$ in the composite Hilbert space
${\cal H}\otimes {\cal H}'$ results in
\BEA
\label{sar}
\hat{{\cal U}}\, |u_k\rangle\otimes |g_1\rangle =\sum_{\alpha=1}^N
\hat{G}_\alpha |u_k\rangle\otimes |g_\alpha\rangle,
\EEA
where $\hat{G}_\alpha$ defined by (\ref{comp}) are operators living
in the Hilbert space ${\cal H}$.

One notes that due to (\ref{comp})
\BEA
\langle g_1|\otimes\langle u_l|\,
\hat{{\cal U}}
\, |u_k\rangle\otimes |g_1\rangle =\delta_{kl}.
\EEA
This specification of $\hat{{\cal U}}$ is not yet complete.  To
complete the definition of $\hat{{\cal U}}$ in the composite Hilbert
space ${\cal H}\otimes {\cal H}'$ one should define its action on
$|u_k\rangle\otimes |g_\alpha\rangle$ for $\alpha=2,...,N$ in addition
to (\ref{sar}). This will suffice, since $|u_k\rangle\otimes
|g_\alpha\rangle $ for $k=1,...,n$ and $\alpha=1,...,N$ is an
orthonormal base in the composite Hilbert ${\cal H}\otimes {\cal H}'$.
This completion is possible to do and one can do that in many
different ways, because it amounts to completing the orthonormal set
of vectors $\hat{G}_\alpha |u_k\rangle\otimes |g_\alpha\rangle$ to an
orthonormal base in ${\cal H}\otimes {\cal H}'$.  Then the columns (or
equivalently the rows) of $\hat{{\cal U}}$ in the base
$|u_k\rangle\otimes |g_\alpha\rangle $ will be a set of of $Nn$
orthonormal vectors, which is equivalent for $\hat{{\cal U}}$ being a
unitary matrix. For a given unitary matrix $\hat{{\cal U}}$ there is a
hermitian operator $\hH_{\rm ov}$ in ${\cal H}\otimes {\cal H}'$ such
that $\hat{{\cal U}}=\exp\left[ \frac{it}{\hbar} \,\hH_{\rm ov}\right]$ 
with some time-parameter $t$. Thus, $\hH_{\rm ov}$ can serve as
a Hamiltonian realizing the needed interaction.

Once (\ref{sar}) is valid for the fixed base $\{|u_k\rangle\}_{k=1}^n$,
one gets for an arbitrary $\hrho$
in ${\cal H}$:
\BEA
\hat{{\cal U}}\, 
\hrho\,\hat{{\cal U}}^\dagger=
\sum_{\alpha=1}^N
\hat{G}_\alpha\,\hrho\, \hat{G}_\alpha^\dagger
\otimes |g_\alpha\rangle\langle g_\alpha|.
\label{ssdd}
\EEA

To complete the construction, one now measures in ${\cal H}'$ any
hermitian operator with a non-degenerate spectrum having the base
$\{|g_\alpha\rangle\}_{\alpha=1}^N$ as its eigenbase. Then the POVM
(\ref{comp}, \ref{durman}, \ref{yale}) accounts for what is happening
| after the interaction and after the selective measurement |
with the initial ensemble described by $\hrho$.

\section{Derivation of Eqs.~(\ref{q1}, \ref{q11}).}
\label{bochka}

One notes from (\ref{gibbs}, \ref{unita}):
\BEA
\hU^\dagger_\tau\,\hrho\,\hU_\tau=\frac{\exp{\left[-\beta 
\hU^\dagger_\tau\,\hH\,\hU_\tau\right]}}{Z}
=\frac{e^{-\beta\Del-\beta\hH}}{Z},
\EEA
where we used the definition $\Del=\hU^\dagger_\tau\,\hH\,\hU_\tau-\hH$
of $\Del$.

Note the standard equality
\BEA
\label{kkk}
e^{-\beta\Del -\beta\hH}=\overrightarrow{\exp}\left[
-\int_0^\beta\d s\,e^{-s\hH}\,\Del \,e^{s\hH}
\right]\,e^{-\beta\hH},
\EEA
where $\overrightarrow{\exp}$ means time-antiordered exponent. 
The easiest way to derive Eq.~(\ref{kkk}) is to
differentiate both sides of it over $\beta$, and note that
they both satisfy to the same first-order differential equation with
the same boundary condition at $\beta=0$.

One now gets
\BEA
\hU^\dagger_\tau\,\hrho\,\hU_\tau=
\overrightarrow{\exp}\left[
-\int_0^\beta\d s\,e^{-s\hH}\Del \,e^{s\hH}
\right]\,\hrho.
\EEA
Tracing out both sides, one finally obtains (\ref{q1}, \ref{q11}).

The simplest way to derive Peierls-Bogoliubov inequality from
(\ref{q1}, \ref{q11}) is to note
the well-known extremal feature of the free energy:
\BEA
&&-T\ln\tr e^{-\beta\hH-\beta\Del}\nonumber\\
&&={\it min}
\left\{
\tr[\hrho\,(\hH+\Del)]+T\tr (\hrho\ln\hrho)
\right\},\nonumber
\EEA
where the minimization is taken over all possible density matrices.
This can alternatively be written as
\BEA
&&\tr\, e^{-\beta\hH-\beta\Del}\nonumber\\
&&={\it max}
\exp\left\{
-\beta\tr[\hrho\,(\hH+\Del)]-\tr (\hrho\ln\hrho)
\right\}.
\label{har}
\EEA
The desired inequality is got by putting the particular
density matrix $\hrho=e^{-\beta\hH}/Z$ in the RHS of (\ref{har}).

\section{Proof of Horn's theorem.}
\label{hornappendix}

We intend to prove that given two vectors $x^{\rm T}=(x_1\geq...\geq x_n)$
and $y^{\rm T}=(y_1\geq...\geq y_n)$ with the following majorization 
relation
\BEA
\label{krot1}
x\prec y
\EEA
there is a real orthogonal matrix $O=(O_{ij})$ such that
\BEA
x_i=\sum_jO_{ij}^2 y_j\,\Leftrightarrow\,
x={\rm diag}\left[\,O\,{\rm diag}[y]\,O^{\rm T}\,
\right].
\label{krot3}
\EEA
Here ${\rm diag}[y]$ means the $n\times n$ diagonal matrix formed by
the vector $y$, ${\rm diag}\left[\,O\,{\rm diag}[y]\,O^{\rm
T}\,\right]$ is the vector formed by diagonal elements of the matrix
$O\,{\rm diag}[y]\,O^{\rm T}$, and $O^{\rm T}$ means transposition:
$\,(O^{\rm T})_{kl}=O_{lk}$. Note that for any orthogonal matrix
$O_{ij}$, the matrix $O_{ij}^2$ is always double-stochastic:
$\sum_{i}O_{ij}^2=\sum_jO_{ij}^2=1$, while the converse is not true
\cite{major1}.

For a given orthogonal matrix $O_{ij}$ there are many unitary matrices
$Q_{ij}$ such that $O^2_{ij}=|Q_{ij}|^2$; e.g.,
$Q_{ij}=e^{i\phi_j}O_{ij}$, where $\phi_j$ are arbitrary phases.

The proof is adopted from Ref.~\cite{ando} and
will be realized in two steps.

\subsection{First step.}

First, one shows that (\ref{krot1}) implies
\BEA
\label{krot2}
x=T_1\,T_2...\,T_{n-1}\,y,
\EEA
where matrices $T$ are so-called T-transform defined as follows.  Any
T-transform $T(m,l;t)$ has three parameters: $m<l$ and $t$, where $m$
and $l$ are natural numbers between zero and $n$, and where $0<t<1$. Its
action on any vector $y$, $y^{\rm T}=(y_1,...,y_n)$, is defined as
\BEA
\label{bars1}
z=T(m,l;t)y,
\EEA
where
\BEA
z^{\rm T}=(y_1,..,y_{m-1},ty_m+(1-t)y_l,y_{m+1},...,\nonumber\\
y_{l-1},(1-t)y_m+ty_l,y_{l+1},...,y_n)
\EEA

To get the matrix of $T(m,l;t)$ starting from $n\times n$ unity matrix
$\hat{1}$, one proceeds as follows:
\BEA
&&(\hat{1})_{mm}=1\,\to\,(\,T(m,l;t)\,)_{mm}=t,\nonumber\\
&&(\hat{1})_{ll}=1\,\to\,(\,T(m,l;t)\,)_{ll}=t,\nonumber\\
&&(\hat{1})_{ml}=0\,\to\,(\,T(m,l;t)\,)_{ml}=1-t,\nonumber\\
&&(\hat{1})_{lm}=0\,\to\,(\,T(m,l;t)\,)_{lm}=1-t,
\label{bars2}
\EEA
while all other elements of the unity matrix are left unchanged.

Eq.~(\ref{krot2}) can now be proven by induction. It is obvious for $n=2$.
Assume it holds for $n-1$. As seen from (\ref{krot1}, 
\ref{krab2}, \ref{krab3}), one has $y_n\leq x_1
\leq y_1$, so there exists an index $k$ that 
\BEA
\label{dan}
y_{k}\leq x_1\leq y_1.
\EEA
This implies 
\BEA
x_1=ty_1+(1-t)y_k,
\EEA
for some $0\leq t\leq 1$. Define a T-transform $T(1,k;t)$ via
\BEA
\label{bars77}
\left(\begin{array}{r}
x_1 \\
\\
\bar{y}~ \\
\end{array}\right)=
T(1,k;t)\,y,
\EEA
where
\BEA
\bar{y}^{\rm T}=(y_2,...,y_{k-1},(1-t)y_1+ty_{k},y_{k+1},..,y_n).
\EEA
It is straighforward to show that 
\BEA
\label{bars7}
\bar{y}^{\rm T}\succ (x_2,..,x_n).
\EEA
Since we assumed (\ref{krot1}) $\Rightarrow$
(\ref{krot2}) is valid for $n-1$, there is a product 
of T-transforms such that $(x_2,..,x_n)^{\rm T}=T_2...\,T_{n-1}\,\bar{y}$.

Thus the implication (\ref{krot1}) $\Rightarrow$ (\ref{krot2}) is proven
by induction.

\subsection{Second step.}

Let us finally prove the implication (\ref{krot1}) $\Rightarrow$
(\ref{krot3}).  Note that for any T-transform $T(m,l;t)$ one can
associate an orthogonal matrix $V(m,l;t)$ by reshaping 
(\ref{bars2}) as follows:
\BEA
&&(\hat{1})_{mm}=1\,\to\,(\,V(m,l;t)\,)_{mm}=\sqrt{t},\nonumber\\
&&(\hat{1})_{ll}=1\,\to\,(\,V(m,l;t)\,)_{ll}=\sqrt{t},\nonumber\\
&&(\hat{1})_{ml}=0\,\to\,(\,V(m,l;t)\,)_{ml}=-\sqrt{1-t},\nonumber\\
&&(\hat{1})_{lm}=0\,\to\,(\,V(m,l;t)\,)_{lm}=\sqrt{1-t}.
\label{bars3}
\EEA
Then (\ref{bars1}) is equivalent to
\BEA
z={\rm diag}\left[\,
V(m,l;t)~{\rm diag}[y]~V^{\rm T}(m,l;t)\,
\right].
\EEA
To prove the implication (\ref{krot1}) $\Rightarrow$
(\ref{krot3}) one again proceeds by induction. It is obviously valid
for $n=2$. One assumes its validity for $n-1$. 
This means (\ref{bars7}) can be re-written as:
\BEA
(x_2,...,x_n)^{\rm T}={\rm diag}\left[\,
\tilde{V}~{\rm diag}[\bar{y}]~\tilde{V}^{\rm T}\,
\right],
\EEA
where $\tilde{V}$ is some orthogonal matrix. To complete the proof,
define an orthogonal matrix
\BEA 
O=\left(\begin{array}{rr}
1 & 0 \\
0 & \tilde{V} \\
\end{array}\right)\,V,
\EEA
where the matrix $V$ corresponds to the T-transform $T$ 
defined in (\ref{bars77}) ( via the correspondence described
in (\ref{bars3})), and rewrite (\ref{bars7}, \ref{bars77}) as
\BEA
x={\rm diag}\left[\,
O~{\rm diag}[\bar{y}]~O^{\rm T}\,
\right].
\label{tt}
\EEA 
This proves the implication (\ref{krot1}) $\Rightarrow$ (\ref{krot3}).

\subsection{An example.}

Let us realize explicitly the above construction for an example:
\BEA
x=(0,0,0),\qquad y=(2,1,-3).
\EEA
For the index $k$ and for the parameter $t$ mentioned before (\ref{dan})
one has
\BEA
k=3,\qquad t=\frac{3}{5}.
\EEA
Analogously:
\BEA
\left(\begin{array}{r}
0 \\
\\
0 \\
\\
0
\end{array}\right)=
\left(\begin{array}{rrr}
1 & 0 & 0 \\
\\
0 & \frac{1}{2} & \frac{1}{2} \\
\\
0 & \frac{1}{2} & \frac{1}{2}\\
\end{array}\right)
\left(\begin{array}{rrr}
\frac{3}{5} & 0 & \frac{2}{5} \\
\\
0 & 1 & 0 \\
\\
\frac{2}{5} & 0 & \frac{3}{5}\\
\end{array}\right)
\left(\begin{array}{r}
2 \\
\\
1 \\
\\
-3
\end{array}\right)
\EEA
realizes the relation between $x$ and $y$ via T-transforms.
Finally the orthogonal matrix $O$ in (\ref{tt}) 
reads for the present example:
\BEA
O=\left(\begin{array}{rrr}
\sqrt{\frac{3}{5}} & 0 & -\sqrt{\frac{2}{5}} \\
\\
-\sqrt{\frac{1}{5}} & \sqrt{\frac{1}{2}} & -\sqrt{\frac{3}{10}} \\
\\
\sqrt{\frac{1}{5}} & \sqrt{\frac{1}{2}} & \sqrt{\frac{3}{10}}\\
\end{array}\right).
\EEA

\section{Derivation of Eq.~(\ref{dard1}).}
\label{barbi}

Here we find the maximum of
\BEA
\label{koza}
\langle w^2\rangle
=\sum_{\alpha=1}^N\lambda_\alpha w^2_\alpha
=\sum_{\alpha=1}^N
\frac{
\widetilde{\langle\psi_\alpha|}\Del |
\widetilde{\psi_\alpha\rangle}^2}
{\widetilde{\langle\psi_\alpha|}
\widetilde{\psi_\alpha\rangle}},
\EEA
where the maximization is taken over all possible decompositions
(\ref{baba}) of the mixed state $\hrho$ into pure states. Using 
(\ref{chibukh1}, \ref{lotos}) one writes equivalently
\BEA
\langle w^2\rangle
=\sum_{\alpha=1}^N\frac{\left[\,
\tr\, (\Y\, \hPi_\alpha)
\,\right]^2}
{\tr\, \hrho\hPi_\alpha},
\label{olen}
\EEA
where
\BEA
\label{sobaka}
\hPi_\alpha=|\pi_\alpha\rangle\langle\pi_\alpha|.
\EEA

The maximization in (\ref{olen}) is taken over all decompositions
of unity
\BEA
\sum_{\alpha=1}^N \hPi_\alpha=\hat{1},
\EEA
where operators $\hPi_\alpha$ live
in the $n$-dimensional Hilbert space ${\cal H}$.

The general idea of the following method was adopted from~\cite{caves}.
Introduce an operator $\hX$ via
\BEA
\label{dub}
\Y=\Re (\hrho \hX)\equiv
\half \left(
\hrho\hX+\hX\hrho
\right),
\EEA
then
\BEA
\tr (\Y\hPi_\alpha)=\Re\,\tr (\hPi_\alpha \hrho \hX).
\EEA
Recall Cauchy inequality
\BEA
|\tr (\hat{A}\hat{B})|^2\leq \tr (\hat{A}\hat{A}^\dagger) \,\tr
(\hat{B}\hat{B}^\dagger), 
\label{cauchy}
\EEA
which is valid for any operators $\hat{A}$
and $\hat{B}$, with the equality being realized for
\BEA
\hat{A}=\alpha\hat{B}^\dagger, 
\EEA
where $\alpha$ is a number.

Applying (\ref{dub}, \ref{cauchy}):
\BEA
\label{bereza1}
\left[\,\tr\, (\Y\, \hPi_\alpha)
\,\right]^2\leq |\tr (\hPi_\alpha \hrho \hX)|^2\\
\equiv |\tr (
\hPi^{1/2}_\alpha\,\hrho^{1/2}\,\hrho^{1/2}\,\hX\,
\hPi^{1/2}_\alpha
)|^2\nonumber\\
\leq \tr (\hPi_\alpha\,\hrho)~
\tr (\hrho\,\hX\,\hPi_\alpha\,\hX),
\label{bereza2}
\EEA
one gets for (\ref{olen}, \ref{sobaka}):
\BEA
\label{zver}
\sum_{\alpha=1}^N
\frac{\left[\,
\tr\, (\Y\, \hPi_\alpha)
\,\right]^2}
{\tr\, \hrho\hPi_\alpha}
\leq \sum_{\alpha=1}^N
\tr (\hrho\,\hX\,\hPi_\alpha\,\hX)\\
=\tr (\hrho\hX^2)=\tr (\Y \hX)
\label{zverzver}
\EEA

Eq.~(\ref{bereza1}) is realized as equality for
\BEA
\label{koshka1}
\Im\,\tr (\hPi_\alpha \hrho \hX)=0,
\EEA
while the Cauchy inequality (\ref{zver}) becomes equality for 
\BEA
\label{koshka2}
\hrho^{1/2}\,\hX|\pi_\alpha\rangle=a_\alpha
\hrho^{1/2}|\pi_\alpha\rangle,
\EEA
where $a_\alpha$ are some numbers.

Both conditions (\ref{koshka1}, \ref{koshka2}) are realized
simultaneously by taking $\{|\pi_\alpha\rangle\}_{\alpha=1}^n$ 
and $\{a_\alpha\}_{\alpha=1}^n$ as, respectively,
eigenvectors and eigenvalues of the hermitian operator $\hX$.  
The representation
(\ref{dard2}) for $\hX$ follows from (\ref{dub}).  The desired
equation (\ref{dard1}) is seen from (\ref{zver}, \ref{zverzver}).

\section{Derivation of Eq.~(\ref{avo}).}
\label{haar}

Here we calculate the
average $\{\langle w^2\rangle\}_{\rm av}$ of
$\langle w^2\rangle$, given by (\ref{koza}), over the
measure (\ref{haarmeasure}).  Using Eq.~(\ref{kara}) it is
straightforward to see that all the terms in the summation in the RHS
of (\ref{koza}) produce the same average. Thus,
\BEA
\frac{\{\langle w^2\rangle\}_{\rm av}}{N}
=\frac{\int{\cal D}M
\,\delta \left[
\sum_{i=1}^N|M_{i}|^2 -1\right]\phi\{M_i\}
}{\int
{\cal D}M
\,\delta \left[
\sum_{i=1}^N|M_{i}|^2 -1\right]},~~~~~~~~
\EEA
where we denoted
\BEA
{\cal D}M
=\prod_{i=1}^N\,\d \Re M_{i}
\,\d \Im M_{i}
\EEA 
and where one notes from (\ref{kara})
\BEA
\phi\{M_i\}=\left|\sum_{j,k=1}^n
M_jM_k^*\sqrt{p_jp_k}
\langle\ep_k|\Del|\ep_j\rangle
\right|^2.
\EEA
Passing to polar coordinates 
\BEA \int
{\cal D}M
=\int_0^{2\pi}\prod_{i=1}^N\d \varphi_i
\int_0^\infty \prod_{i=1}^N|M_i|\,\d |M_i|
\EEA
one gets
\BEA
\frac{\{\langle w^2\rangle\}_{\rm av}}{N}=
\sum_{j,k=1}^np_jp_k
\langle\ep_j|\Del|\ep_j\rangle
\langle\ep_k|\Del|\ep_k\rangle\,\frac{I_{ij}}{I_0},
\EEA
where
\BEA
&&I_{jk}=\int_0^\infty
\prod_{i=1}^N\d z_i~~
\delta\left[
\sum_{i=1}^Nz_i-1
\right]\,
\frac{z_jz_k}{\sum_{l=1}^np_lz_l},~~\\
&&I_{0}=\int_0^\infty
\prod_{i=1}^N\d z_i~~
\delta\left[
\sum_{i=1}^Nz_i-1
\right].
\EEA

These integrals are calculated for $j,k=1,..,n$
by the same method. For example,
\BEA e^{-r}
r^NI_{jj}=
\int_0^\infty
\prod_{i=1}^N\d y_i~~
\delta\left[
\sum_{i=1}^N y_i-r
\right]\,
\frac{y_j^2\,e^{-r}}{\sum_{l=1}^np_ly_l},\nonumber
\EEA
\BEA
\label{pompilius1}
\Gamma(N+1)I_{jj}&&=
\int_0^\infty
\prod_{i=1}^N\d y_i\,y^2_j~~
\frac{e^{-\sum_{i=1}^N y_i}}{\sum_{l=1}^np_ly_l},\\
\label{pompilius2}
&&=
\int_0^\infty
\prod_{l=1}^n\d y_l\,y^2_j~~
\frac{e^{-\sum_{l=1}^n y_l}}{\sum_{l=1}^np_ly_l},\\
&&=\int_0^\infty\d s
\int_0^\infty
\prod_{i=1}^N\d y_i\,y^2_j~~
e^{-\sum_{l=1}^ny_l(sp_l+1)}\nonumber
\EEA
where when passing from (\ref{pompilius1}) from (\ref{pompilius2})
we changed the integration variable, $z_i=y_i/r$
and integrated over $r$
from $0$ to $\infty$.

Further calculations are straightforward and lead to (\ref{avo}).
For dealing with this equation the following
formula is useful:
\BEA
\int_0^\infty\d s\,
\prod_{k=1}^n
\frac{1}{\theta_m+s}=
\sum_{k=1}^n\ln \theta_k\prod_{l\not =k}^n\frac{1}{\theta_l-\theta_k},~
\label{mumu00}
\EEA
where $\theta_k$'s are some positive numbers.

\section{}
\label{sanik}

Here we give an example of the situation discussed around (\ref{tut},
\ref{nono}). The effect announced there does exist neither for
two-level systems |simply because for a $2\times 2$ traceless matrix
$\Del$ a zero eigenvalue implies $\Del=0$,| nor for three level
systems. This last fact requires some calculations which will be
omitted.

The simplest situation which supports the effect is thus a four-level
system. The following example was inspired by \cite{ozawa}. Consider
a four-level system with Hamiltonian:
\BEA
\label{zzz1}
\hH=
\left(\begin{array}{rr}
\hat{A} & 0\\
0& \hat{B}\\
\end{array}\right),
\EEA
where $\hat{A}$ and $\hat{B}$ are $2\times 2$ matrices:
\BEA
\label{zzz1.5}
\hat{A}=
\left(\begin{array}{rr}
a & b\\
b& d\\
\end{array}\right),\quad
\hat{B}=
\left(\begin{array}{rr}
a & b\\
b& c\\
\end{array}\right),
\EEA
with $a$, $b$, $c$ and $d$ being some real numbers.

Assume that the unitary operator $\hU_\tau$ is given as
an exchange interaction:
\BEA
\hU_\tau=
\left(\begin{array}{rr}
0 & \hat{1} \\
\hat{1} & 0\\
\end{array}\right),
\EEA
where $\hat{1}$ is the
$2\times 2$ unit matrix.

The Hamiltonian $\hU^\dagger_\tau\,\hH\,\hU_\tau$ in the Heisenberg
representation at time $\tau$ then reads
\BEA
\hU^\dagger_\tau\,\hH\,\hU_\tau
=\left(\begin{array}{rr}
\hat{B} & 0\\
0& \hat{A}\\
\end{array}\right).
\label{zzz2}
\EEA
It is seen from the block-diagonal form of $\hH$
and $\hU^\dagger_\tau\,\hH\,\hU_\tau$ that they both
have the same eigenvalues.

As follows from (\ref{zzz1}, \ref{zzz1.5}, \ref{zzz2}),
the matrix $\Del$,
\BEA
\Del=\left(\begin{array}{rrrr}
0 & 0 ~~& 0& 0~~\\
\\
0& c-d & 0& 0~~\\
\\
0& 0~~& 0& 0~~\\
\\
0& 0~~& 0& d-c
\end{array}\right),
\label{bom}
\EEA
has a doubly degenerate eigenvalue equal to zero, and the corresponding
eigenvectors can be taken as
\BEA
|0_1\rangle=
\left(\begin{array}{r}
1\\
0\\
0\\
0\\
\end{array}\right),\quad
|0_2\rangle=
\left(\begin{array}{r}
0\\
0\\
1\\
0\\
\end{array}\right).
\EEA

It is now obvious that though
\BEA
\left\langle 0_1\left| \,\left[
\hU^\dagger_\tau\,\hH\,\hU_\tau
\right]^m
\right| 0_1\right\rangle-
\langle 0_1| \hH^m  | 0_1\rangle
=0,\quad {\rm for}\quad m=1,2,
\nonumber
\EEA

one still has

\BEA
\left\langle 0_1\left| \,\left[
\hU^\dagger_\tau\,\hH\,\hU_\tau
\right]^3
\right| 0_1\right\rangle-
\langle 0_1| \hH^3  | 0_1\rangle
=b^2(c-d)\not=0,\nonumber
\EEA
\BEA
\left\langle 0_1\left| \,\left[
\hU^\dagger_\tau\,\hH\,\hU_\tau
\right]^4
\right| 0_1\right\rangle-
\langle 0_1| \hH^4  | 0_1\rangle
\nonumber\\
=2ab^2(c-d)+b^2(c^2-d^2)\not=0.
\EEA
These relations were used in (\ref{tsakat}).

\end{document}